\documentclass[3p]{elsarticle}  

\journal{Computational Statistics and Data Analysis}

\usepackage{listings}
\usepackage{amsmath,amssymb} 
\usepackage{graphicx}
\usepackage{graphics}
\usepackage{epsfig}
\usepackage{tabularx}
\usepackage{multirow}
\usepackage{amssymb}
\usepackage{multirow}
\usepackage{color}
\usepackage{url}
\usepackage[symbol*]{footmisc}
\usepackage{algorithm}
\usepackage{algorithmic}

\newcommand{\by}{{\bf y}}

\newcommand{\rw}{\rightarrow}

\newcommand{\sS}{{\sf S}}

\newcommand{\mbE}{\mathbb{E}}

\newcommand{\bftheta}{{\boldsymbol{\theta}}}

\newtheorem{theorem}{\bf Theorem}
\newtheorem{Assumption}{\bf A.}

\begin{document}

\begin{frontmatter}



\title{A nonlinear population Monte Carlo scheme for the Bayesian estimation of parameters of $\alpha$-stable distributions}

\author{Eugenia Koblents$^a$\footnote{\textit{E-mail address:} ekoblents@tsc.uc3m.es.}}

\author{Joaqu\'{\i}n M\'{\i}guez$^b$}

\author{Marco A. Rodr\'{\i}guez$^c$}

\author{Alexandra M. Schmidt$^d$}

\address{$^a$Universidad Carlos III de Madrid, Department of Signal Theory and Communications, \\ Avda. de la Universidad 30, 28911 Legan\'es, Spain. \\
$^b$Queen Mary University of London, School of Mathematical Sciences,\\Mile End Rd, E1 4NS London, UK.\\
$^c$Universit\'e du Qu\'ebec \`a Trois-Rivi\`eres, D\'epartement
des Sciences de l'Environnement, \\
C.P. 500 Trois-Rivi\`eres QC G9A 5H7, Qu\'ebec, Canada. \\ 
$^d$Universidade Federal do Rio de Janeiro, Departamento de
M\'etodos Estat\'isticos, \\ Caixa Postal 68530, CEP.: 21945-970, Rio
de Janeiro, Brazil. } 


\begin{abstract}
The class of $\alpha$-stable distributions enjoys multiple practical applications in signal processing, finance, biology and other areas because it allows to describe interesting and complex data patterns, such as asymmetry or heavy tails, in contrast with the simpler and widely used Gaussian distribution. The density associated with a general $\alpha$-stable distribution cannot be obtained in closed form, which hinders the process of estimating its parameters. A nonlinear population Monte Carlo (NPMC) scheme is applied in order to approximate the posterior probability distribution of the parameters of an $\alpha$-stable random variable given a set of random realizations of the latter. The approximate posterior distribution is computed by way of an iterative algorithm and it consists of a collection of samples in the parameter space with associated nonlinearly-transformed importance weights. 
A numerical comparison of the main existing methods to estimate the $\alpha$-stable parameters is provided, including the traditional frequentist techniques as well as a Markov chain Monte Carlo (MCMC) and a likelihood-free Bayesian approach. It is shown by means of computer simulations that the NPMC method outperforms the existing techniques in terms of parameter estimation error and failure rate for the whole range of values of $\alpha$, including the smaller values for which most existing methods fail to work properly. Furthermore, it is shown that accurate parameter estimates can often be computed based on a low number of observations. Additionally, numerical results based on a set of real fish displacement data are provided.
\end{abstract}
\end{frontmatter}

\section{Introduction}
\label{Introduction}

\subsection{$\alpha$-stable distributions}

The family of $\alpha$-stable distributions \cite{Nolan2013} is a rich class of probability distributions
that displays many patterns of shapes, allowing for
asymmetry and heavy tails, opposite to the widely used, but more restrictive, Gaussian distribution. The class of $\alpha$-stable distributions has
been found suitable for statistical modelling in signal processing,
finance and biology, among other fields \cite{Nolan2013}. For this reason,
efficient computational algorithms for the estimation of the parameters of
$\alpha$-stable distributions in practical setups are needed.

A random variable is stable if a linear combination of two independent copies of the
variable has the same distribution, up to location and scale
parameters. An $\alpha$-stable distribution is a generalization of
the Gaussian distribution and stems from a more general version of
the central limit theorem, avoiding the assumption of finite
variance \cite{Nolan2013}.

We denote a general $\alpha$-stable distribution as $\mathcal{S}(\alpha, \beta,
\gamma, \delta)$, where $\alpha \in (0,2]$ is a stability index (or characteristic exponent),
$\beta \in [-1,1]$ is a skewness parameter, and $\gamma > 0$ and
$\delta \in \mathbb{R}$ determine the scale and location, respectively. The ``shape'' of
the distribution is determined by $\alpha$ and $\beta$: lower values of
$\alpha$ correspond to heavier tails and a sharper peak, while
$\beta$ determines the degree and sign of asymmetry ($\beta > 0$
corresponding to right-skewness). As $\alpha \rightarrow 2$, the
distribution approaches a (non-standard) Gaussian distribution,
and $\beta$ becomes less meaningful and harder to estimate
accurately. As $\alpha \rightarrow 0$, the effect of $\beta$ becomes
more pronounced, the density gets extremely high at the peak and the tails become heavier.
Stable distributions have one single tail for $\alpha < 1$ and
$\beta = \pm 1$, and both tails otherwise. The mean of an $\alpha$-stable distribution is only defined if $1 < \alpha \leq 2$ and the variance is only finite for $\alpha = 2$ (Gaussian case).

Distributions of the $\alpha$-stable class have several specific mathematical properties. All (non-degenerate) stable distributions are unimodal, continuous and have an infinitely differentiable probability density function (pdf) \cite{Nolan2013}. However, the pdf is not available in closed form except for a few particular cases (Gaussian, Cauchy and L\'evy) \cite{Nolan2013}, a fact that has hampered a broader use of stable distributions in practice. The $\alpha$-stable distribution is generally specified in terms of its characteristic function $\Phi(u) = E [\exp (iuX)]$, where $E[\cdot]$ denotes expectation, $X$ is the random variable and $i = \sqrt{-1}$. In this work we consider the so called 0-parameterization \cite{Nolan2013} of the characteristic function
\begin{equation*}
\Phi(u) = \left\{ \begin{array}{ll}
\exp\left\{ i\delta u - \gamma^\alpha |u|^\alpha \left[ 1 + i\beta \tan(\frac{\pi \alpha}{2}) \textrm{sign}(u) (|\gamma u|^{(1-\alpha)}-1) \right] \right\}, & \textrm{if} \; \alpha \neq 1 \\
\exp\left\{ i\delta u - \gamma |u| \left[ 1 + i\beta \frac{2}{\pi}
\textrm{sign}(u) \log(\gamma |u|) \right] \right\}, & \textrm{if} \;
\alpha = 1
                    \end{array}
 \right..
\end{equation*}
This parameterization is continuous in all the parameters and is more
suitable for numerical work and statistical inference than
alternative representations that can be found in the literature
\cite{Nolan2013}. It has to be noted that, under the 0-parameterization, the scale parameter is not the standard deviation (even in the Gaussian case). On the other hand, for $1 < \alpha \leq 2$ the mean is of the form $\delta_0 - \beta \gamma(\tan \frac{\pi \alpha}{2})$, where $\delta_0$ denotes the location parameter in the 0-parameterization \cite{Nolan2013}.



\subsection{Parameter estimation}

A large number of methods for the estimation of the parameters of $\alpha$-stable distributions have been proposed in the last decades, since the initial work of \cite{Fama1971}. However, accurate estimation of all four parameters, especially when $\alpha$ is low, is still an open problem and an active area of research. The difficulty of evaluating the pdf associated to an $\alpha$-stable distribution (except for a few particular cases), as well as the posterior dependencies among the parameters, make the parameter estimation problem hard. 

A computationally simple method based on data sample quantiles and look-up tables was proposed in \cite{Mcculloch1986}, as a generalization of the method in \cite{Fama1971}, but it is known to yield consistent parameter estimates only when $0.4 \le \alpha \le 2$. In \cite{Nolan2013} a modified quantile method is proposed which is claimed to work for any values of the parameters. It allows to estimate low values of $\alpha$, but yields poorer estimates of $\beta$ than the standard quantile method. In \cite{Koutrouvelis1981}, an iterative weighted regression procedure was introduced that fits the parameters to the empirical characteristic function (ECF) estimated from the data. This technique 
does not provide solutions for low values of $\alpha$ either. In \cite{Kogon1998} a simplified and improved version of the method in \cite{Koutrouvelis1981} is proposed which greatly reduces the amount of computation by restricting the estimation to an interval of the characteristic function. In \cite{Nolan2001} a maximum likelihood approach was proposed based on a numerical evaluation of the likelihood \cite{Nolan1997}. This method uses the quantile estimator of \cite{Mcculloch1986} as an initial approximation to the parameters and maximizes the likelihood via an approximate gradient based search. It implements a fast likelihood evaluation but its use is restricted to cases when $\alpha > 0.4$. The fractional lower order moments and the log absolute moments methods have been proposed in \cite{Nikias1995} for the symmetric case ($\beta = \delta = 0)$. Both methods are computationally simple but the latter has proved to be more efficient in practice \cite{Nikias1995}. Extensions of these methods have been proposed for the asymmetric case with $\delta = 0$ in \cite{Kuruoglu2001}. These modified methods require transformations of the data into symmetrized and centered sequences, reducing the available sample size in one half and two thirds, respectively.  When the amount of data is small, as considered here, this results in numerical problems and inconsistent estimates.

In the Bayesian framework, several attempts have been made to estimate the parameters of $\alpha$-stable distributions by using Markov chain Monte Carlo (MCMC) algorithms \cite{Gilks2005,Robert04}. In \cite{Buckle1995} a Gibbs sampler is proposed, which requires sampling from a high-dimensional auxiliary variable and has, therefore, a high computational cost. The random walk Metropolis-Hastings (MH) sampler proposed in \cite{Lombardi2007} relies on a likelihood approximation using the inverse fast Fourier transform (FFT) \cite{Menn2006} of the characteristic function near the mode and Bergstr\"{o}m expansions for the tails. This sampler is very sensitive to the value of $\alpha$, which determines the threshold between these two regions, as well as the spacing between the FFT samples. For this reason, it is very hard to tune the algorithm in such a way that acceptable results can be guaranteed for any $\alpha$.

Likelihood-free or approximate Bayesian computation (ABC) methods have also been applied to this problem. This family of algorithms avoids the evaluation of the likelihood function using forward simulation from the observation model \cite{Turner2012,DelMoral2012}. The MCMC and ABC approaches can be combined into MCMC-ABC methods \cite{Marjoram2003}, that explore the parameter space iteratively using Markov chains with the desired stationary distribution. These techniques are particularly likely to get stuck or yield extremely high rejection rates, requiring a prohibitive computational cost even for simple problems. The partial rejection control (PRC)-ABC algorithm was developed in \cite{Sisson2007} as an alternative to MCMC-ABC methods, which suffer from severe mixing problems. In \cite{Peters2012} the PRC-ABC algorithm was applied to the Bayesian inference problem in univariate and multivariate $\alpha$-stable models. In this paper we focus on the population Monte Carlo (PMC)-ABC algorithm proposed in \cite{Beaumont2009}, which has been shown to yield the best performance among the existing ABC methods \cite{Turner2012}.

In this work, we are especially interested in the particular case when the $\alpha$ parameter is very low ($\alpha < 0.5$), when most of the existing techniques fail to perform properly. Data with such properties can arise in the biological sciences or engineering, where the deviation with respect to average behaviour can be very large \cite{Belanger2001,Niranjayan2010}. In particular, in this paper we are interested in problems where only a small set of heavy-tailed observations are available, which is the case in many practical applications \cite{Belanger2001}. To the best of our knowledge this problem has not been specifically addressed in the literature, probably because of the lack of adequate inference methods.





\subsection{Contributions and organization of the paper}

In this work we propose to apply a nonlinear population Monte Carlo (NPMC) method \cite{Koblents2013a} to the problem of estimating the parameters of $\alpha$-stable distributions. The NPMC algorithm is an iterative importance sampling (IS) scheme that computes nonlinearly transformed weights to mitigate the degeneracy problem common to conventional IS methods \cite{Bengtsson08,Koblents2013a}. We resort to an approximation of the likelihood function based on the method proposed in \cite{Nolan1997} in order to compute the importance weights. This approximation introduces an (additional) distortion of the weights that is not accounted for in the analysis of \cite{Koblents2013a}. We address this issue and provide an explicit error bound for the estimates produced by an importance sampler with nonlinearly transformed approximate weights.
In addition to these theoretical results, we provide computer simulations that show that the NPMC algorithm outperforms the main existing methods for any value of the parameter $\alpha$ in the interval $(0,2]$. Moreover, the NPMC scheme has a low computational cost compared to other Bayesian schemes (e.g., \cite{Lombardi2007,Peters2012}) and our simulations show that it can be robust in data-poor scenarios. 
Finally, we present numerical results for a set of real data, corresponding to daily displacements of a set of fish in Ganelon Creek (Canada) in 1998. This dataset is demanding because only a small number of observations per individual ($\approx 30$) is available and the data present extremely heavy tails for many individuals.

The remaining of this paper is organized as follows. 
The proposed NPMC inference algorithm is described in Section \ref{NPMC_alg}. Exhaustive computer simulations that illustrate the performance of the NPMC method 
as well as some of the main existing methods, based on synthetic data, are presented and discussed in Section \ref{Sims}. Simulation results obtained with a set of real 
fish displacement data are shown in Section \ref{Sims_peces}.
Finally, Section \ref{Conclusions} is devoted to the conclusions.

\section{Algorithm}
\label{NPMC_alg}

\subsection{Bayesian inference in $\alpha$-stable models}

Let $\boldsymbol{\theta} = [\alpha, \beta, \gamma,
\delta]^\top$ be a vector containing the parameters of an
$\alpha$-stable distribution and let $\textbf{y} = [y_{1},
\ldots, y_{T}]^\top$ be a vector of $T$ independent and identically
distributed (i.i.d.) samples from $\mathcal{S}(\alpha,
\beta, \gamma, \delta)$. We adopt a Bayesian approach and aim at approximating
the posterior probability distribution of $\boldsymbol{\theta}$
given the observation vector $\textbf{y}$ using a Monte Carlo
scheme. The density associated to the posterior distribution is
denoted $p(\boldsymbol{\theta} | \textbf{y}) \propto p(\textbf{y} |
\boldsymbol{\theta}) p(\boldsymbol{\theta})$, where
$p(\boldsymbol{\theta})$ and $p(\textbf{y} | \boldsymbol{\theta})$
are the prior distribution and the likelihood function of the
parameters $\boldsymbol{\theta}$, respectively. The likelihood
function is factorized as $p(\textbf{y} | \boldsymbol{\theta}) = \prod_{t=1}^{T}
\mathcal{S}(y_{t} ; \alpha, \beta, \gamma, \delta)$, where $\mathcal{S}(y_t; \alpha, \beta, \gamma, \delta)$ denotes the pdf of the $\alpha$-stable distribution. Since this density does not have a closed form, it cannot be evaluated exactly and a numerical approximation is needed. While other possibilities exist, we adopt the approximation proposed in \cite{Nolan1997}.

\subsection{Nonlinear population Monte Carlo method}

The population Monte Carlo (PMC) method \cite{Cappe04} is an iterative IS scheme that seeks to approximate a target probability distribution with associated pdf $\pi(\boldsymbol{\theta})$. The PMC method generates a sequence of proposal pdfs $q_\ell(\boldsymbol{\theta})$, $\ell = 1, \ldots, L$, that is expected to yield increasingly better approximations of the target pdf as the algorithm converges. However, IS-based techniques, including PMC methods, suffer from the well known weight degeneracy problem \cite{Bengtsson08,Koblents2013a}. This problem arises when the proposal pdf is not well-tailored to the target pdf, yielding extreme variations of the importance weights (IWs) and a very low sampling efficiency. Weight degeneracy hinders the application of IS and PMC techniques in many practical applications, in favor of the family of MCMC techniques.

The effort in the field of PMC algorithms has been typically directed toward the design of efficient proposal functions \cite{Cappe08,Djuric11}. Alternatively, in \cite{Koblents2013a} a nonlinear PMC (NPMC) scheme was proposed that specifically addresses the weight degeneracy problem. The emphasis is not placed on the proposal update scheme, which can be very simple. The main feature of the technique is the application of a nonlinear transformation to the IWs in order to reduce their variations. In this way, the efficiency of the sampling scheme is improved (especially when drawing from poor proposals) and the degeneracy of the IWs is drastically mitigated even when the number of generated samples is relatively small. In this work we apply the \emph{clipping} transformation of the IWs described in \cite{Koblents2013a}. The proposed algorithm is displayed in Algorithm \ref{alg:NPMC_algo}.

\begin{algorithm}[ht]
\caption{Nonlinear PMC targeting $\pi (\boldsymbol{\theta}) = p(\boldsymbol{\theta} | \textbf{y})$.} \label{alg:NPMC_algo}
\begin{algorithmic}
\REQUIRE ($\ell = 1, \ldots, L$):
\begin{enumerate}

\item Draw $M$ samples $\{
\boldsymbol{\theta}_\ell^{(i)} \}_{i=1}^M$ from the proposal density $q_\ell(\boldsymbol{\theta})$:

\begin{itemize}
\item at iteration $\ell=1$, let $q_1 (\boldsymbol{\theta}) = p(\boldsymbol{\theta})$, where the prior distribution $p(\boldsymbol{\theta})$ is assumed to have a support set $\mathcal{S} \in \mathbb{R}^4$.

\item at iterations $\ell=2,\ldots,L$, $q_\ell(\boldsymbol{\theta})$ is the truncated Gaussian approximation of $p(\boldsymbol{\theta} | \textbf{y})$ with support $\mathcal{S}$ obtained at iteration $\ell-1$.

\end{itemize}

\item For $i=1,\ldots,M$, compute the unnormalized IWs
\begin{equation*}
w_\ell^{(i)*} \propto \frac{\hat{p} ( \textbf{y} |
\boldsymbol{\theta}_\ell^{(i)} ) p ( \boldsymbol{\theta}_\ell^{(i)}
) }{q_\ell ( \boldsymbol{\theta}_\ell^{(i)} )},
\end{equation*}
where $\hat{p} ( \textbf{y} | \boldsymbol{\theta}_\ell^{(i)} )$ denotes the likelihood approximation computed using the method of \cite{Nolan1997}.

\item For $i=1,\ldots,M$, compute TIWs by \emph{clipping} the IWs as
$\bar{w}_\ell^{(i)*} = \min ( w_\ell^{(i)*}, \mathcal{T}_\ell^{M_T}
)$ and normalize them as $\bar{w}_\ell^{(i)} = \bar{w}_\ell^{(i)*}/ \sum_{j=1}^M \bar{w}_\ell^{(j)*}$.
The threshold value $\mathcal{T}_\ell^{M_T}$ denotes the $M_T$-th
highest unnormalized IW $w_\ell^{(i)*}$, with $1 < M_T < M$.

\item Construct a truncated Gaussian approximation $q_{\ell+1} (\boldsymbol{\theta}) = \mathcal{TN} (\boldsymbol{\theta};
\boldsymbol{\mu}_\ell, \boldsymbol{\Sigma}_\ell)$ of the posterior
$p(\boldsymbol{\theta} | \textbf{y})$, with support $\mathcal{S}$ and mean vector and
covariance matrix computed as
\begin{equation}
\label{eq_mu_sigma}
\boldsymbol{\mu}_{\ell} = \sum_{i=1}^M \bar{w}_\ell^{(i)}
\boldsymbol{\theta}_{\ell}^{(i)}, \quad
\boldsymbol{\Sigma}_{\ell} = \sum_{i=1}^M \bar{w}_\ell^{(i)} (
\boldsymbol{\theta}_{\ell}^{(i)} - \boldsymbol{\mu}_{\ell}
)( \boldsymbol{\theta}_{\ell}^{(i)} -
\boldsymbol{\mu}_{\ell} )^\top.
\end{equation}
\end{enumerate}
\end{algorithmic}
\end{algorithm}

In this paper we consider a proper uniform prior pdf for all the parameters, defined on a restricted support $\mathcal{S} \in \mathbb{R}^4$, and construct a truncated multivariate Gaussian proposal pdf $q_\ell(\bftheta)$ at each iteration $\ell=2,\ldots,L$, with the same support as the prior pdf, based on the set of previous samples and IWs. This basic proposal construction can be improved in various ways, yielding slightly better results in some cases. For example, a defensive approach can be applied \cite{Cappe08}, in which a subset of samples are generated from the prior pdf at each iteration, to ensure that the support of the proposal pdf contains that of the target pdf. This can be useful in those cases when the posterior pdf of some parameters presents heavier tails than a Gaussian pdf. Heavy-tailed proposals can also be constructed as multivariate Cauchy or Student's \textit{t} distributions. Alternatively, the proposal pdf can be built as a continuous approximation of the target pdf using the samples generated by the NPMC algorithm and a smoothing kernel, at the expense of an increased computational cost.

The likelihood approximation $\hat{p}(\textbf{y} | \boldsymbol{\theta}_\ell^{(i)})$ required in step 2 is computed following \cite{Nolan1997}, where an accurate method is provided to compute general stable densities and distribution functions for essentially all values of the parameters. The method in \cite{Nolan1997} is implemented in Nolan's STABLE program (available in the website \cite{WebNolan}) and in Mark Veillette's Matlab function \textsf{stblpdf}, publicly available as part of the toolbox \textsf{stbl} in the website \cite{WebVeillette}. This toolbox uses an alternative parameterization of the characteristic function. Thus, a translation of the location parameter $\delta$ is needed in order to use this function under the 0-parameterization. A discussion on the accuracy of Nolan's STABLE toolbox is provided in \cite{Gentle2012}.

For the \textit{clipping} procedure performed in step 3 we consider, at each iteration $\ell$, a permutation $i_1, \ldots, i_M$ of the indices in $\{1, ..., M \}$ such that $w_\ell^{(i_1)*} \geq \ldots
\geq w_\ell^{(i_M)*}$ and choose a \textit{clipping} parameter $M_T < M$. We select the threshold value $\mathcal{T}_\ell^{M_T} := w_\ell^{(i_{M_T})*}$ and apply \textit{clipping} to the largest IWs, i.e., 
$\bar{w}_\ell^{(i_k)*} = \mathcal{T}_\ell^{M_T}$ for $k=1, \ldots, M_T$, while $\bar{w}_\ell^{(i_k)*} = w_\ell^{(i_k)*}$ for $k = M_T+1, \ldots, M$. This transformation leads to flat TIWs in the region of interest of $\boldsymbol{\theta}$, yielding a baseline of $M_T$ \textit{effective} samples (those with non-negligible weights) at each iteration, allowing for a robust update of the proposal pdf.

The proposed NPMC method can also be applied in cases where the observations are not independent, as long as the likelihood function can be evaluated or well approximated. For example, in \cite{Koblents2013a} the proposed method has been applied to estimate the parameters and hidden states in state-space models.

\subsection{Asymptotic convergence of NIS with approximate weights}
\label{Analysis}


The clipping of the IWs in step 3 of the NPMC algorithm introduces a distortion in the random probability measure generated by a nonlinear importance sampler and, therefore, it is not apparent, a priori, that this measure should converge in the same way as the measure induced by the standard IWs. This issue is addressed in \cite{Koblents2013a}, where it is shown that, as long as $\frac{M_T}{M}\rw 0$, the approximation of integrals of bounded functions using IWs and using TIWs both converge to the same value. However, the analysis in \cite{Koblents2013a} only yields error rates for convergence in probability and, more importantly, it relies on the ability to compute the IWs exactly --which is not the case for the problem addressed in this paper. 

In this section we look explicitly into the convergence of the estimates of integrals computed using approximate weights. In particular, we provide upper bounds for the estimation errors that hold almost surely (a.s.) and depend explicitly on both the number of generated samples, $M$, and the approximation error for the IWs. The same as in \cite{Koblents2013a}, the analysis is valid for a single-stage importance sampler; we do not incorporate the iterations (over the index $\ell$) of the NPMC algorithm.


Recall that $\pi(\bftheta) = p(\bftheta|{\bf y})$ is the target pdf, let $q(\bftheta)$ be the importance function used to generate samples in an IS scheme (not necessarily normalized) and let $h(\bftheta) \propto \pi(\bftheta)$ be a function proportional to $\pi$, with the proportionality constant independent of $\bftheta$. The samples drawn from $q$ are $\bftheta^{(i)}$, $i=1, ..., M$, and their associated non-normalized IWs are 
$w^{(i)*} = h( \bftheta^{(i)} ) / q ( \bftheta^{(i)})$, $i=1, ..., M$.

We introduce the weight function $g=h/q$, hence $g( {\bftheta}^{(i)}) = w^{(i)*}$. The support of $g$ is the same as the support of $q$ and $\pi$, denoted $\sS \subseteq \mathbb{R}^K$. If we assume that both $q(\bftheta)>0$ and $\pi(\bftheta) > 0$ for any $\bftheta \in \sS$, then $g(\bftheta) > 0$ for every $\bftheta \in \sS$ as well (a standard assumption in classical IS). It is also apparent that $\pi \propto gq$, with the proportionality constant independent of $\bftheta$. 

If the IWs can be computed exactly, the approximation $\pi^M$ of $\pi$ can be
written as
\begin{equation*}
\pi^M(d\bftheta) = \sum_{i=1}^M w^{(i)}
\delta_{{\bftheta}^{(i)}}(d\bftheta),
\end{equation*}
where $w^{(i)} = \frac{
    g ( {\bftheta}^{(i)} )
}{
    \sum_{j=1}^M g ( {\bftheta}^{(j)} )
}$, $i=1, ..., M$.

If the weight function can only be computed approximately, let us denote its approximation as $g^\epsilon$. The resulting random measure is
\begin{equation*}
\pi^{M,\epsilon}(d\bftheta) = \sum_{i=1}^M w^{(i),\epsilon}
\delta_{{\bftheta}^{(i)}}(d\bftheta),
\end{equation*}
where $w^{(i),\epsilon} = \frac{
    g^\epsilon ( {\bftheta}^{(i)} )
}{
    \sum_{j=1}^M g^\epsilon ( {\bftheta}^{(j)} )
}$, $i=1, ..., M$. 

Let us denote by $\varphi^M$ the clipping transformation used to compute non-normalized TIWs, i.e., $\bar{w}^{(i)*} = \varphi^M ( w^{(i)*} )$, $i=1,\ldots, M$. The weighted approximation of $\pi(\bftheta)d\bftheta$ constructed according to the nonlinear IS scheme is
\begin{equation*}
\bar \pi^{M,\epsilon}(d\bftheta) = \sum_{i=1}^M \bar w^{(i),\epsilon}
\delta_{{\bftheta}^{(i)}}(d\bftheta),
\end{equation*}
where $\bar w^{(i),\epsilon} = \frac{\varphi^M ( g^\epsilon ( {\bftheta}^{(i)}
))}{\sum_{j=1}^M \varphi^M ( g^\epsilon ( {\bftheta}^{(j)}))} $, $i=1, ...,
M$.


We make the following assumptions on the weight function, $g$, and its approximation, $g^\epsilon$.
\begin{Assumption} \label{asApprox_g}
For any $\epsilon \ge 0$, the approximation $g^\epsilon$ of the weight function satisfies the inequality
$$
\sup_{\bftheta \in \sS} | g(\bftheta) - g^\epsilon(\bftheta) | \le \epsilon \quad \mbox{a.s.}
$$
\end{Assumption}
\begin{Assumption} \label{asBounds_on_g}
The weight function $g$ has a finite upper bound and a positive lower bound. Specifically, there exists a real number $0 < a < \infty$ such that $ a^{-1} \le g(\bftheta) \le a $ for every $\bftheta \in \sS$.
\end{Assumption}
\begin{Assumption} \label{asBounds_on_geps}
The same bounds of $g$ hold for its approximations $g^\epsilon$, $\epsilon \ge 0$. To be specific, $a^{-1} \le g^J(\bftheta) \le a$ for every $\bftheta \in \sS$ and any $\epsilon \ge 0$.
\end{Assumption}

Note that if the support set $\sS$ is compact then A.\ref{asBounds_on_g} holds whenever $q>0$ and $h>0$ in $\sS$. Otherwise, the proposal $q$ has to be chosen so that it has heavier tails than $\pi$.

In the sequel we look into the approximation of integrals of the form 
$$
(f,\pi) = \int I_{\sS}(\bftheta) f(\bftheta) \pi(\bftheta) d\bftheta,
$$ 
where $I_\sS(\bftheta)$ is an indicator function (namely, $I_\sS(\bftheta)=1$ if $\bftheta \in \sS$ and $I_\sS(\bftheta)=0$ otherwise) and $f$ is a bounded real function in the parameter space $\sS$. We use $\|f\|_\infty =
\sup_{\bftheta \in \sS}|f(\bftheta)| < \infty$ to denote the supremum norm of a bounded function. The set of bounded functions on $\sS$ is $B(\sS) = \{ f: \sS \rw \mathbb{R} : \|f\|_\infty < \infty\}$. The approximations of interest are
$$
(f,\pi^{M,\epsilon}) = \sum_{i=1}^M f(\bftheta^{(i)}) w^{(i),\epsilon}
\quad \mbox{and} \quad
(f,\bar\pi^{M,\epsilon}) = \sum_{i=1}^M f(\bftheta^{(i)}) \bar w^{(i),\epsilon}.
$$

The following Theorem yields upper bounds for the absolute approximation errors $| (f,\pi^{M,\epsilon}) - (f,\pi) |$ and $| (f,\bar \pi^{M,\epsilon}) - (f,\pi) |$ that depend explicitly on $M$ and $\epsilon$.

\begin{theorem} \label{thBasic}
Assume that $M_T \le \sqrt{M}$ and assumptions A.\ref{asApprox_g}, A.\ref{asBounds_on_g} and A.\ref{asBounds_on_geps} hold. Then, there exist positive and a.s. finite random variables $W_{f,\upsilon}$ and $\bar W_{f,\upsilon}$, independent of $M$ and $\epsilon$, such that
\begin{equation}
| (f,\pi^{M,\epsilon}) - (f,\pi) | \le \frac{
    W_{f,\upsilon}
}{
    M^{\frac{1}{2}-\upsilon}
} + C \epsilon \label{eq1Basic}
\end{equation}
and
\begin{equation}
| (f,\bar \pi^{M,\epsilon}) - (f,\pi) | \le \frac{
    \bar W_{f,\upsilon}
}{
    M^{\frac{1}{2}-\upsilon}
} + C \epsilon \label{eq2Basic}
\end{equation}
for every $f \in B(\sS)$, arbitrarily small $0<\upsilon<\frac{1}{2}$ and some finite constant $C$. Both $C$ and $\upsilon$ are independent of $M$, $M_T$ and $\epsilon$.
\end{theorem}

A proof is provided in \ref{apThBasic}. Theorem \ref{thBasic} yields an upper bound for the (random) absolute error that consists of two terms, one that depends on the number of samples $M$ and another one that depends on the weight approximation error $\epsilon$. As $M \rw \infty$, the first term vanishes with the usual Monte Carlo rate of convergence despite the approximation of the IWs and the clipping transformation. The second term is proportional to the approximation error, hence it only vanishes when the routine used to compute $g^\epsilon$ can be made arbitrarily accurate (i.e., $\epsilon \rw 0$), typically by increasing the computational effort invested in this calculation.

\section{Computer simulations}
\label{Sims}

In this section we provide extensive simulation results to illustrate the performance of the main existing methods for the estimation of $\alpha$-stable parameters. The numerical results are obtained for a set of synthetic observations from $\alpha$-stable distributions with a wide range of parameters. First we consider the NPMC and two other Bayesian methods: an MCMC algorithm and an ABC technique. The implemented NPMC and MCMC methods use the likelihood approximation proposed in \cite{Nolan1997}, while the ABC method is based on a likelihood-free approach. Finally, we compare the NPMC method with the more relevant frequentist methods proposed in the literature. 

\subsection{Performance of the NPMC algorithm}
\label{Sims_NPMC}

We have performed $5000$ independent simulations of the NPMC algorithm to approximate $p(\boldsymbol{\theta} | \textbf{y})$ with different parameter and observation vectors. In
each simulation run, we draw the parameters
$\boldsymbol{\theta} = [\alpha, \beta, \gamma, \delta]^\top$
from a  distribution $\mu(\boldsymbol{\theta}) = \mu(\alpha) \mu(\beta) \mu(\gamma) \mu(\delta)$ constructed from a set of independent uniform components, i.e.,
\begin{equation*}
\mu(\alpha) = \mathcal{U} (\alpha; (0,2]), \; \mu(\beta) =
\mathcal{U} (\beta; [-1,1]), \; \mu(\gamma) = \mathcal{U} (\gamma; (0,10]) \; \textrm{and} \; \mu(\delta) = \mathcal{U} (\delta;
[-5,5]).
\end{equation*}

In each simulation run, we generate a set of $T=30$ samples $y_t$, $t=1,\ldots,T$, from the resulting $\alpha$-stable distribution $\mathcal{S}(\alpha, \beta, \gamma, \delta)$. It is straightforward to generate samples from an $\alpha$-stable distribution using an extension of the Box-M\"uller algorithm \cite{Chambers1976}, which is detailed in \ref{sampling_method} for the 0-parameterization. We have selected such a low number of observations in order to reproduce as closely as possible the setup of the fish displacement dataset studied in Section \ref{Sims_peces}, where around 30 observations are provided for each individual \cite{Belanger2001}. The gathering of the data can be a costly process in real applications, for example in biology, leading to a low number of available observations. As will be shown in the simulations, even such a low number of observations usually allows to accurately identify the parameters. However, the estimation precision is highly sensitive to the actual parameter values. See the supplementary material for a brief simulation study of the performance with larger observation sets.

We consider two different prior distributions for the inference algorithm. On the one hand, we consider a prior distribution $p_1(\boldsymbol{\theta}) = \mu (\boldsymbol{\theta})$. In this case, the support of the prior pdf is $\mathcal{S}_1 = (0,2] \times [-1,1] \times (0,10] \times [-5,5]$. Additionally, we consider a broader prior distribution for $\gamma$ and $\delta$, namely, $p_2(\gamma) = \mathcal{U} (\gamma; (0,100])$ and $p_2(\delta) = \mathcal{U} (\delta; [-50,50])$, to show the algorithm dependence on the prior distribution. Therefore, the support of the prior pdf $p_2$ is $\mathcal{S}_2 = (0,2] \times [-1,1] \times (0,100] \times [-50,50]$. We run the NPMC algorithm with $L=10$ iterations in both settings, with $M = 300$ and $M_T=20$ with the $p_1$ prior, and $M=10^3$ and $M_T = 30$ with the $p_2$ prior.

At each iteration of the NPMC algorithm, $\ell = 1, \ldots, L$, we compute the mean square error (MSE) associated to each parameter $\theta_k$, $k=1,\ldots,4$, as $MSE_{\ell,k} = \sum_{i=1}^M \bar{w}_\ell^{(i)}(\theta_{\ell,k}^{(i)} - \theta_k)^2 = ( \mu_{\ell,k} - \theta_k )^2 +
\sigma_{\ell,k}^2$, where $\mu_{\ell,k}$ is the $k$-th component of the mean vector $\boldsymbol{\mu}_\ell$ and the variance term $\sigma_{\ell,k}^2$ is the $(k,k)$ component of matrix $\boldsymbol{\Sigma}_\ell$ (see Algorithm \ref{alg:NPMC_algo}). The global MSE is obtained as $MSE_\ell = \frac{1}{4} \sum_{k=1}^4 MSE_{\ell,k}$, by averaging over the parameter vector components $\theta_k$, $k=1,\ldots,4$. We additionally compute at each iteration an approximation of the normalized effective sample size (NESS) as $M_\ell^{neff} = [M \sum_{i=1}^M (\bar{w}_\ell^{(i)})^2 ]^{-1}$ \cite{Koblents2013a}, which serves as an indicator of the numerical stability of the algorithm.

\begin{figure*}[ht]
\centering
\includegraphics[width=0.49\textwidth]{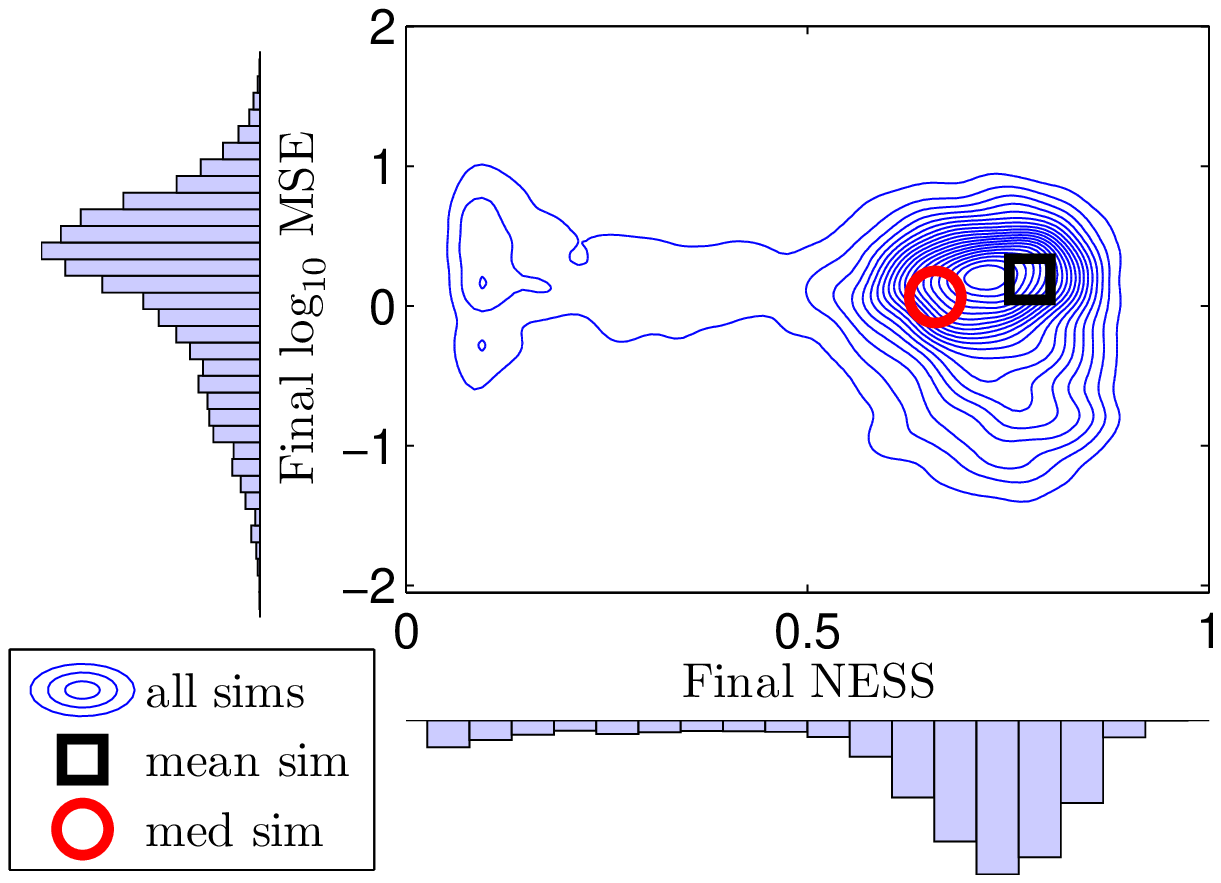}
\includegraphics[width=0.49\textwidth]{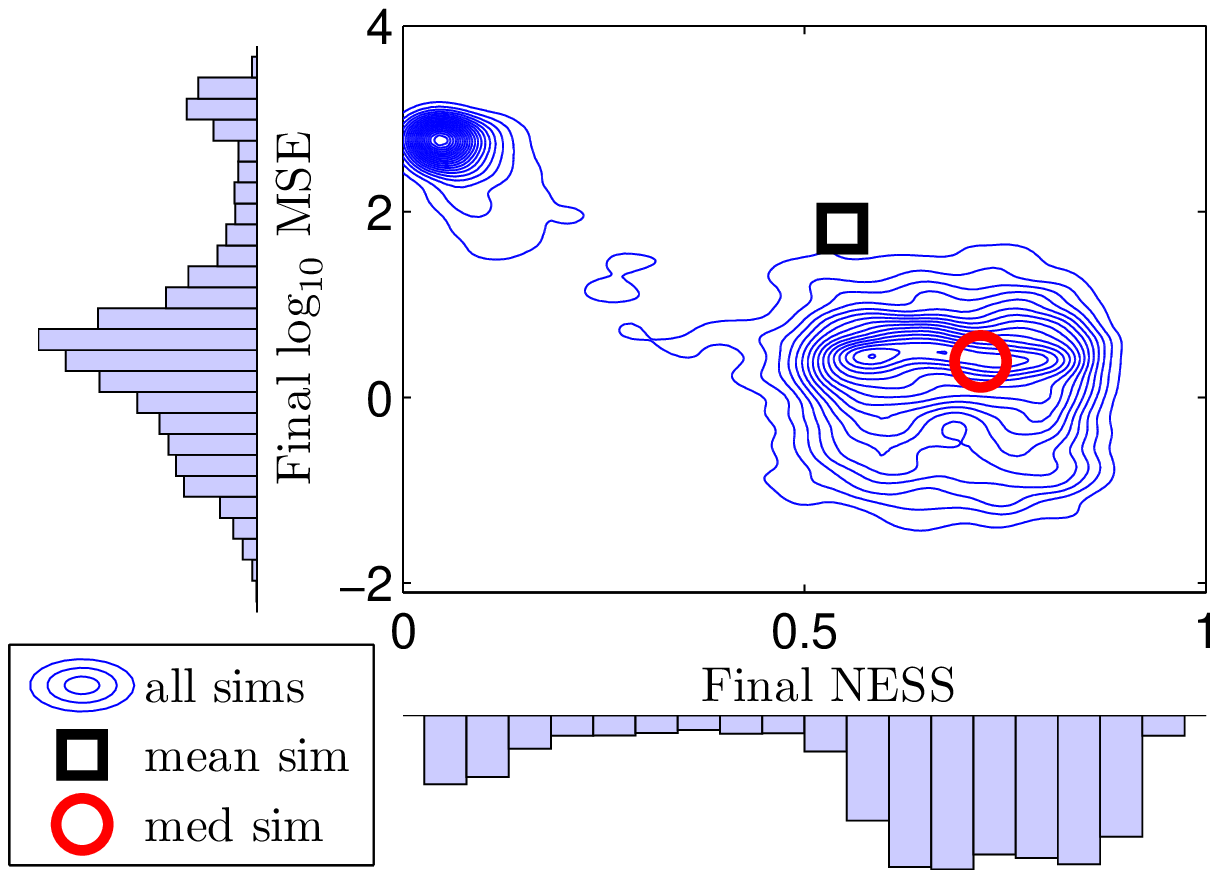}
\vspace{-0.4cm}\caption{Smooth representation of the final MSE versus the final NESS
obtained by the NPMC algorithm in each simulation run, obtained with the $p_1$ (\textit{left}) and $p_2$ (\textit{right}) priors. Average and median simulation runs are depicted with big
squares and circles, respectively.}
\label{fig_1}
\end{figure*}

In Figure \ref{fig_1} a smooth representation of the final MSE values ($MSE_{L}$) 
versus the final NESS ($M_L^{neff}$) values obtained in each of the
$5000$ simulation runs is shown. Results obtained with the narrow prior distribution $p_1(\boldsymbol{\theta})$ (\textit{left}) and the broad prior distribution $p_2(\boldsymbol{\theta})$ (\textit{right}) are displayed. A Gaussian kernel has been used to smooth the discrete sample representations.
The big squares and circles represent simulation runs with a final $MSE_L$ close to the global mean and median values, respectively. As can be observed from the figure, in both cases the final NESS presents bimodality. A subset of the simulations ends up with a low NESS value, yielding higher MSE values on average. When the broader prior is used, we obtain poorer performance, with a low final NESS. However, when the final NESS is $M_L^{neff} > 0.3$, the performance is similar with both choices of the prior. These different behaviors are due to the value of parameter $\alpha$, as will be made clear in the rest of this section.

\begin{figure*}[ht]
\centering
\includegraphics[width=0.49\textwidth]{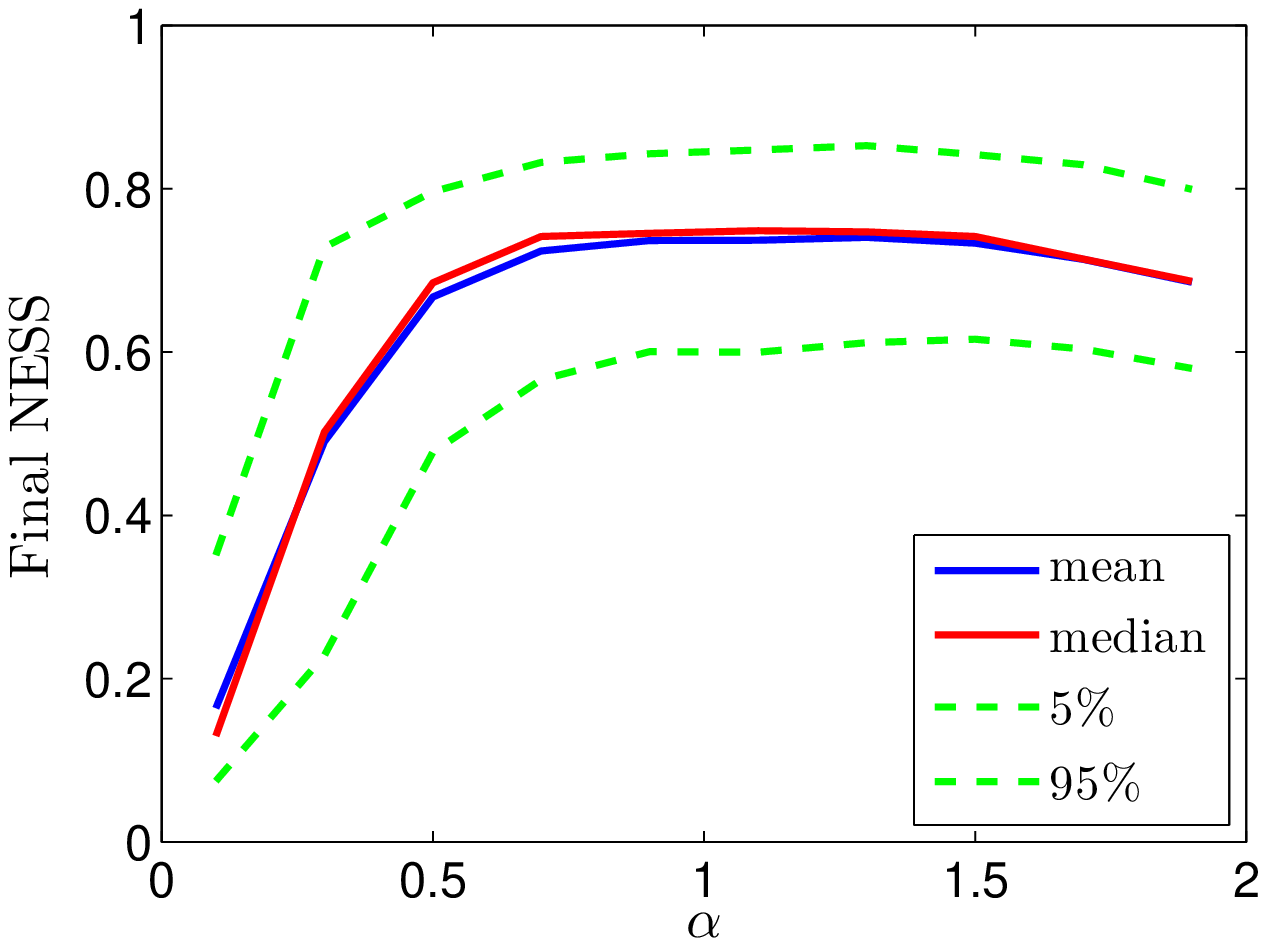}
\includegraphics[width=0.49\textwidth]{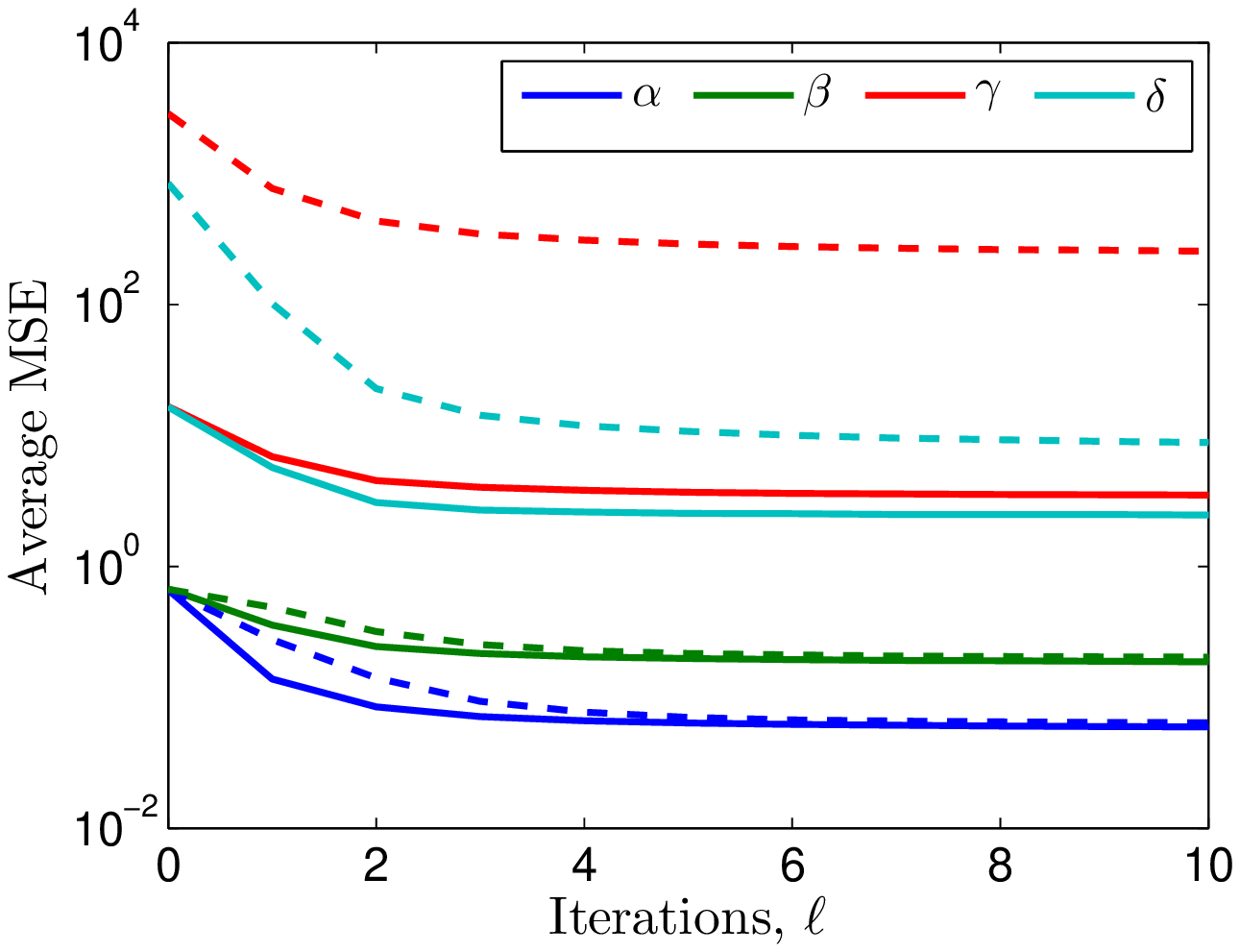}
\vspace{-0.4cm}\caption{\textit{Left}: Final NESS statistics versus the true value of $\alpha$ with the narrow prior pdf $p_1(\boldsymbol{\theta})$. \textit{Right}: Evolution along the iterations of the MSE of each parameter, obtained with the narrow prior $p_1$ (solid lines) and broad prior $p_2$ (dashed lines).} 
\label{fig_2}
\end{figure*}

In Figure \ref{fig_2} (\textit{left}) some statistics (mean, median, 5\% and 95\% quantiles) of the final NESS value are represented versus the true value of $\alpha$. The curves have been obtained from the final NESS values obtained at each simulation run, averaged over intervals of $\alpha$ of length 0.2. It can be observed that low $\alpha$ values (that is, stable distributions with heavy tails) yield low NESS values after convergence of the algorithm. Very similar NESS results are obtained with the broader prior $p_2$.
In Figure \ref{fig_2} (\textit{right}) the evolution along the iterations of the MSE of  each parameter ($MSE_{\ell,k}$) is represented, averaged over $5000$ simulations runs, for the narrow (solid lines) and broad (dashed lines) prior distributions. The initial values $MSE_{0,k}$ have been obtained from the samples drawn from the prior $p(\boldsymbol{\theta})$ at the first iteration, before computing the IWs. It can be observed that the MSE smoothly decreases, reaching a steady value in 5 or 6 iterations. Parameters $\alpha$ and $\beta$ attain similar performance with both choices of the prior, since the corresponding marginal priors are the same under $p_1$ and $p_2$. However, parameters $\gamma$ and $\delta$ attain a significantly poorer performance with the broader prior $p_2$. Especially, the $\gamma$ parameter is estimated more poorly with the $p_2$ prior, on average. 

We have also performed computer simulations with different proposal constructions (not shown). In particular, we have considered a defensive approach, in which 10\% of the samples are generated from the prior pdf at each iteration $\ell > 1$. In this case, slightly better results have been obtained in some cases. On the contrary, using a multivariate Cauchy as a proposal pdf yields a somewhat higher MSE, especially when $\alpha$ is low. For simplicity, in the following we abide by the plain truncated Gaussian proposal pdf.

\subsection{Performance of a Metropolis-Hastings random walk}
\label{MH_section}

In this section we consider a MH algorithm which, similarly to the NPMC method, uses the likelihood approximation proposed in \cite{Nolan1997}. Initially, we have implemented the MH method proposed in \cite{Lombardi2007}, which uses a likelihood approximation based on the inverse FFT of the characteristic function and Bergstr\"{o}m expansions for the tails. However, this algorithm has turned out to be extremely sensitive to the selection of certain key parameters, such as the spacing between the FFT samples or the threshold between the two regions. Thus, it is fairly complicated, if not impossible, to adjust those parameters for a general case, particularly when the distribution of interest is heavy-tailed, as already noted in \cite{Peters2012}. For this reason, we do not present simulation results for the algorithm of \cite{Lombardi2007} in this paper. We have not considered the Gibbs sampling method proposed in \cite{Buckle1995} either since it turns out even more computionally demanding than the random walk MCMC algorithm, as stated in \cite{Lombardi2007}. 

Alternatively, and provided that the method in \cite{Nolan1997} yields a good approximation to the likelihood for almost all values of the parameters (except for $\alpha \approx 0$), we consider a standard MH algorithm which uses the likelihood approximation in \cite{Nolan1997} to compute the acceptance ratio. The Metropolis-Hastings algorithm with the likelihood approximation computed with the method in \cite{Nolan1997} is displayed in \ref{MCMC_Nolan}.



We consider a proposal distribution $q(\bftheta | \bftheta^{(i-1)}) = \mathcal{TN} (\bftheta; \bftheta^{(i-1)}, \boldsymbol{\Sigma})$ constructed as a truncated Gaussian random walk with covariance matrix $\boldsymbol{\Sigma} = \textrm{diag} (0.25, 0.25, 1, 1)$, with the same support $\mathcal{S}$ as the prior pdf. The total chain length has been set to $I=3000$ and $I=10^4$ for the priors $p_1 (\bftheta)$ and $p_2 (\bftheta)$, respectively. This yields a total amount of processed samples equal to that of the NPMC method in Section \ref{Sims_NPMC}. The bulk of the execution time of both techniques is the evaluation of the likelihood approximation for each sample $\bftheta^{(i)}$, and thus both have a very similar computational complexity. The Markov chains generated by the MH algorithm have been post-processed, removing a burn-in period of 10\% of the number of samples $I$ and then thinning by a factor of 9. Thus, we have obtained final sample sets of length $M=300$ and $M=10^4$ for the priors $p_1$ and $p_2$, respectively, the same as for the NPMC method. We have performed 5000 independent simulations with the same settings as the NPMC algorithm in Section \ref{Sims_NPMC}.

\begin{figure*}[ht]
\centering
\includegraphics[width=0.49\textwidth]{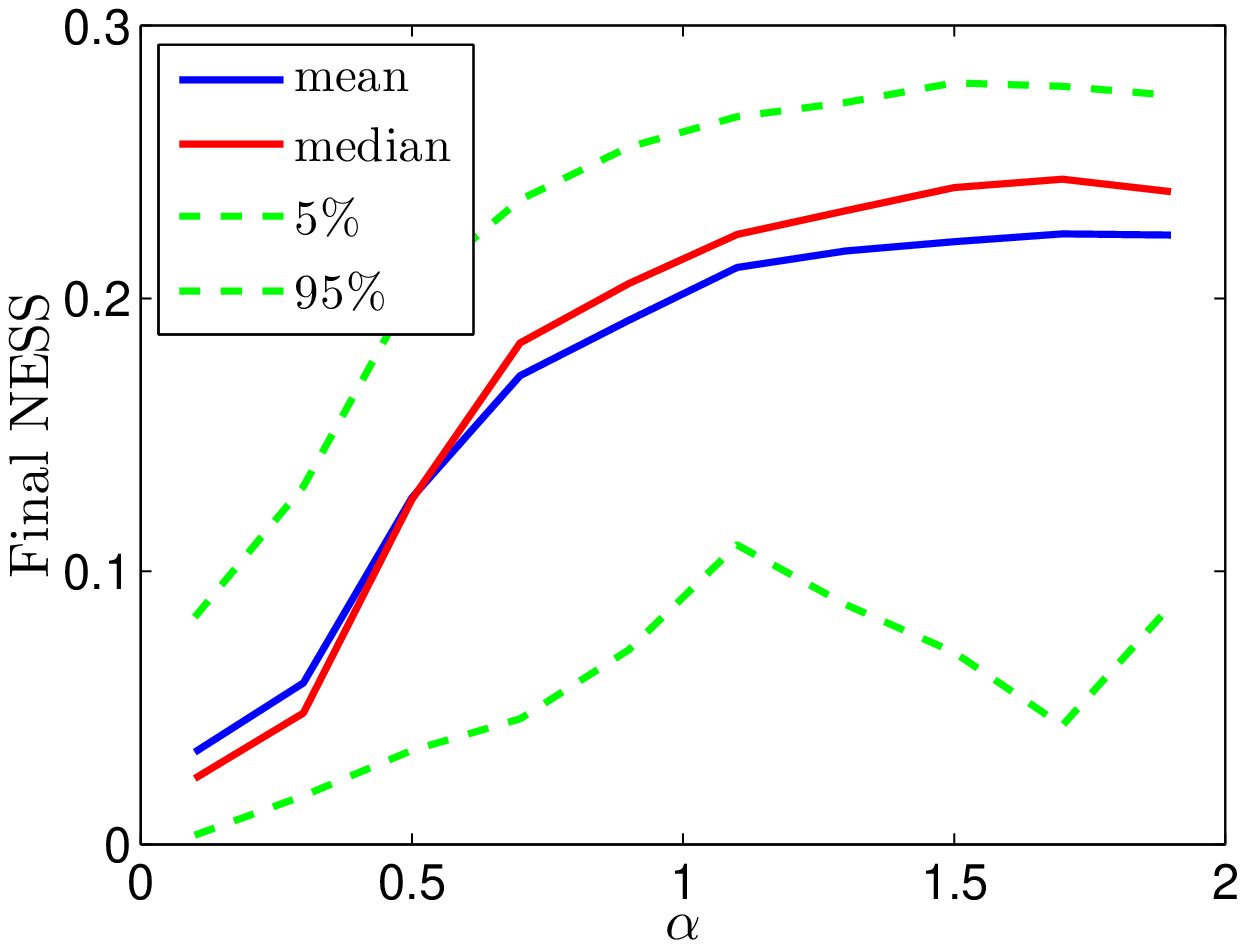}
\includegraphics[width=0.49\textwidth]{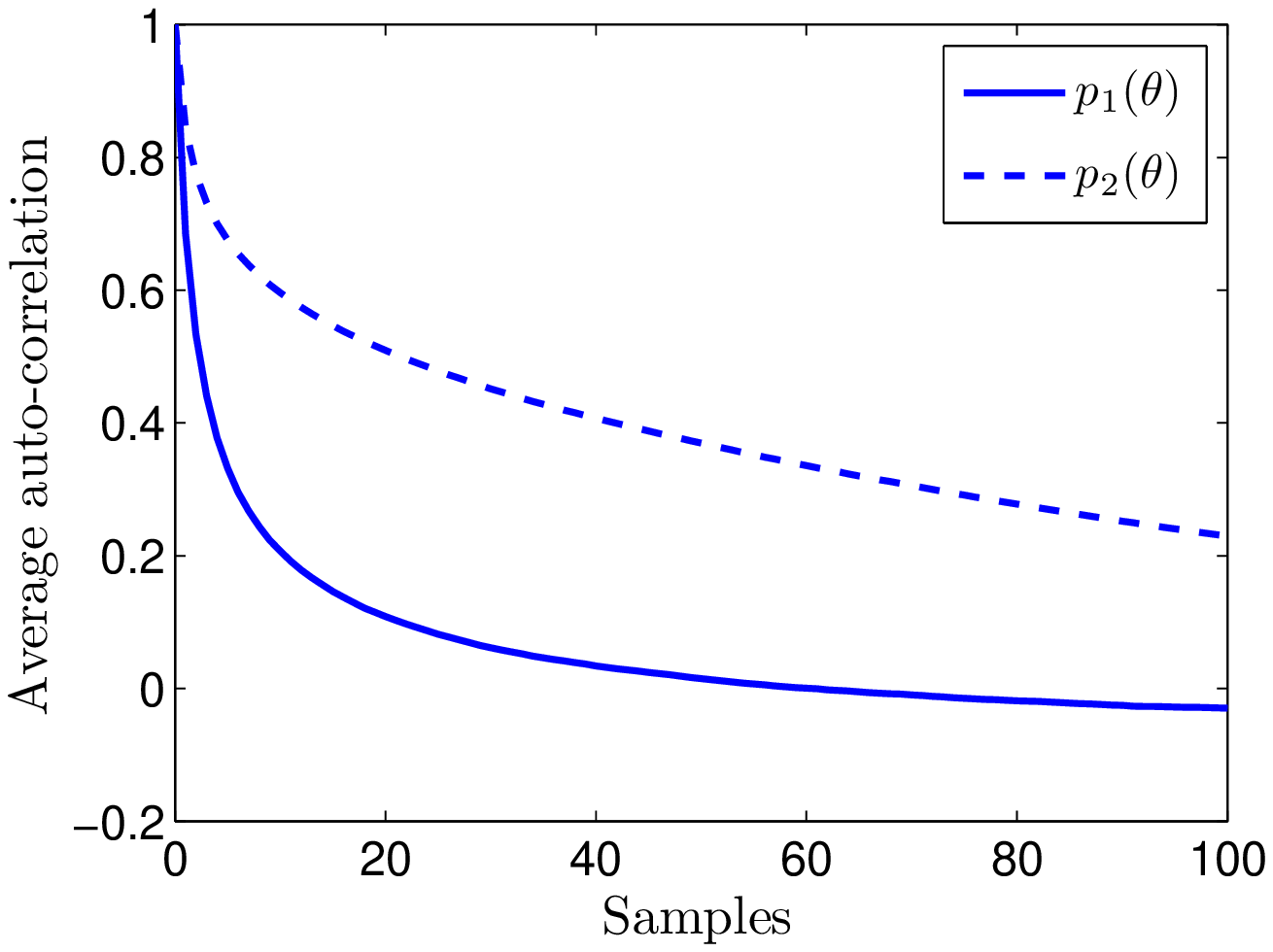}
\caption{\textit{Left}: NESS statistics versus the true value of $\alpha$ obtained by the MH algorithm with the prior distribution $p_1(\bftheta)$. \textit{Right}: Average auto-correlation function of the final chains generated by the MH method, after removing the burn-in period and thinning the output, obtained with priors $p_1(\bftheta)$ and $p_2(\bftheta)$.}
\label{fig_3}
\end{figure*}

In Figure \ref{fig_3} (\textit{left}) statistics of the final average NESS obtained by the MH algorithm are represented versus the true value of $\alpha$, for the prior distribution $p_1(\bftheta)$. Note that in the MCMC literature the ESS is defined differently from that used in IS techniques. In this case, it is an indicator of the size of a i.i.d. sample with the same variance as the current one, and is computed as $M^{neff} = [1+2\sum_{j=1}^\infty \hat{\varrho}(j)]^{-1}$, where $\hat{\varrho}(j) = \textrm{corr}(\bftheta^{(0)}, \bftheta^{(j)})$ is the average autocorrelation function at lag $j$. It can be observed that, similarly to the results obtained with the NPMC method, for low values of $\alpha$ the performance of the algorithm is poorer. In this case, however, even for $\alpha$ values between 1 and 2, the NESS is around 20\%, which indicates that the resulting samples are highly correlated. Figure \ref{fig_3} (\textit{right}) displays the average autocorrelation function obtained from the final Markov chains when either $p_1(\bftheta)$ or $p_2(\bftheta)$ are used as priors. It can be seen that the final samples obtained with the prior $p_2$ present a much higher correlation than with $p_1$ (the final sample size is also larger in this case).

\subsection{Performance of a likelihood-free method}
\label{ABC_section}

The PRC-ABC method was developed in \cite{Sisson2007} as an alternative to MCMC-ABC methods, which suffer from severe mixing problems. In \cite{Peters2012} a PRC-ABC method is applied to the $\alpha$-stable parameters problem, which is claimed to outperform previous Bayesian attempts, such as the Gibbs sampler in \cite{Buckle1995} and the MH method in \cite{Lombardi2007}. However, the PRC-ABC method described in \cite{Peters2012} requires the setting of a large number of parameters, which affect the performance of the method and are very difficult to adjust for arbitrary $\alpha$. We have performed simulations of the PRC-ABC method with the set of parameters suggested by the authors (with the summary statistics computed from the McCulloch quantiles). However, we have obtained highly inaccurate results for most values of $\alpha$. The likelihood-free approximation is claimed to improve as the tolerance level $\epsilon$ decreases, but in practice it becomes inconsistent for low $\epsilon$. As a stopping criterion, the authors propose to run 10 replicate sampler implementations, and to stop the algorithm when the NESS consistently drops below a given threshold, which results in a great increase in the computational complexity. 

On the other hand, a PMC-ABC method was proposed in \cite{Beaumont2009}, as an alternative to the PRC-ABC technique, which has been shown to introduce a bias in the approximation of the posterior. A comparison including the main ABC methods is provided in \cite{Turner2012}, which suggests that the PMC based scheme is the one with the best performance. We have come to the same conclusions through our simulations and, for this reason, we include the PMC-ABC scheme in this comparison, instead of the PRC-ABC method of \cite{Peters2012}. However, we have selected some of the parameters as suggested by \cite{Peters2012}.

The PMC-ABC method performs iterative importance sampling with $L$ iterations, substituting the evaluation of the likelihood function by the ABC approximation based on forward simulations from the observation model. A sequence of decreasing tolerance parameters $\epsilon_1 \geq \ldots \geq \epsilon_L$ has to be specified. At the first iteration, $\ell=1$, the proposal distribution $q_1(\bftheta)$ is selected as the prior, and for iterations $\ell =2, \ldots, L$ it is constructed as a truncated multivariate Gaussian pdf $q_\ell(\bftheta) = \mathcal{TN} (\bftheta; \boldsymbol{\mu}_{\ell-1}, \boldsymbol{\Sigma}_{\ell-1})$, in the same manner as for the NPMC method. At each iteration $\ell = 1,\ldots, L$, pairs of samples $\bftheta_\ell^{(i)} \sim q_\ell(\bftheta)$ and $\by_\ell^{(i)} \sim p(\by | \bftheta_\ell^{(i)})$ are drawn until $M$ samples are accepted. The acceptance rule is that a distance metric $\rho$ between the observations $\by$ and the samples $\by_\ell^{(i)}$ is below a threshold $\epsilon_\ell$, i.e., $\rho (\by, \by_\ell^{(i)}) < \epsilon_\ell$. The metric $\rho$ (the Euclidean distance in our case) is typically computed in terms of some nearly-sufficient low-dimensional summary statistics of the data. At each iteration, the IWs are computed as $w_\ell^{(i)} \propto p(\bftheta_\ell^{(i)}) / q_\ell(\bftheta_\ell^{(i)})$, $i=1,\ldots,M$, which assumes a flat likelihood for the accepted samples and zero likelihood for the rest \cite{Beaumont2009,Turner2012}. The PMC-ABC algorithm is outlined in \ref{PMC_ABC}.

We consider as summary statistics the quantile method estimates of \cite{Mcculloch1986}, as suggested in \cite{Peters2012}. The tolerance parameter sequence has been set to $\epsilon_\ell \in \{ 100, 99, \ldots, 2,1, 0.9, \ldots, 0.1 \}$, with $L=109$ iterations, and the number of samples per iteration has been set to $M=1000$.

The acceptance rate becomes extremely low as the threshold parameter $\epsilon_\ell$ decreases and, particularly, when $\alpha$ is low, which results in a high running time for the algorithm. For this reason, we have limited the execution of this method to 15 minutes (which is far more than the time required by the NPMC and the MH methods to converge under the same setting). We have performed 2500 independent simulations of this algorithm, with the prior distribution $p_1(\bftheta)$ only. Around 50\% of the simulations reached iteration $\ell = 100$.

\begin{figure*}[ht]
\centering
\includegraphics[width=0.49\textwidth]{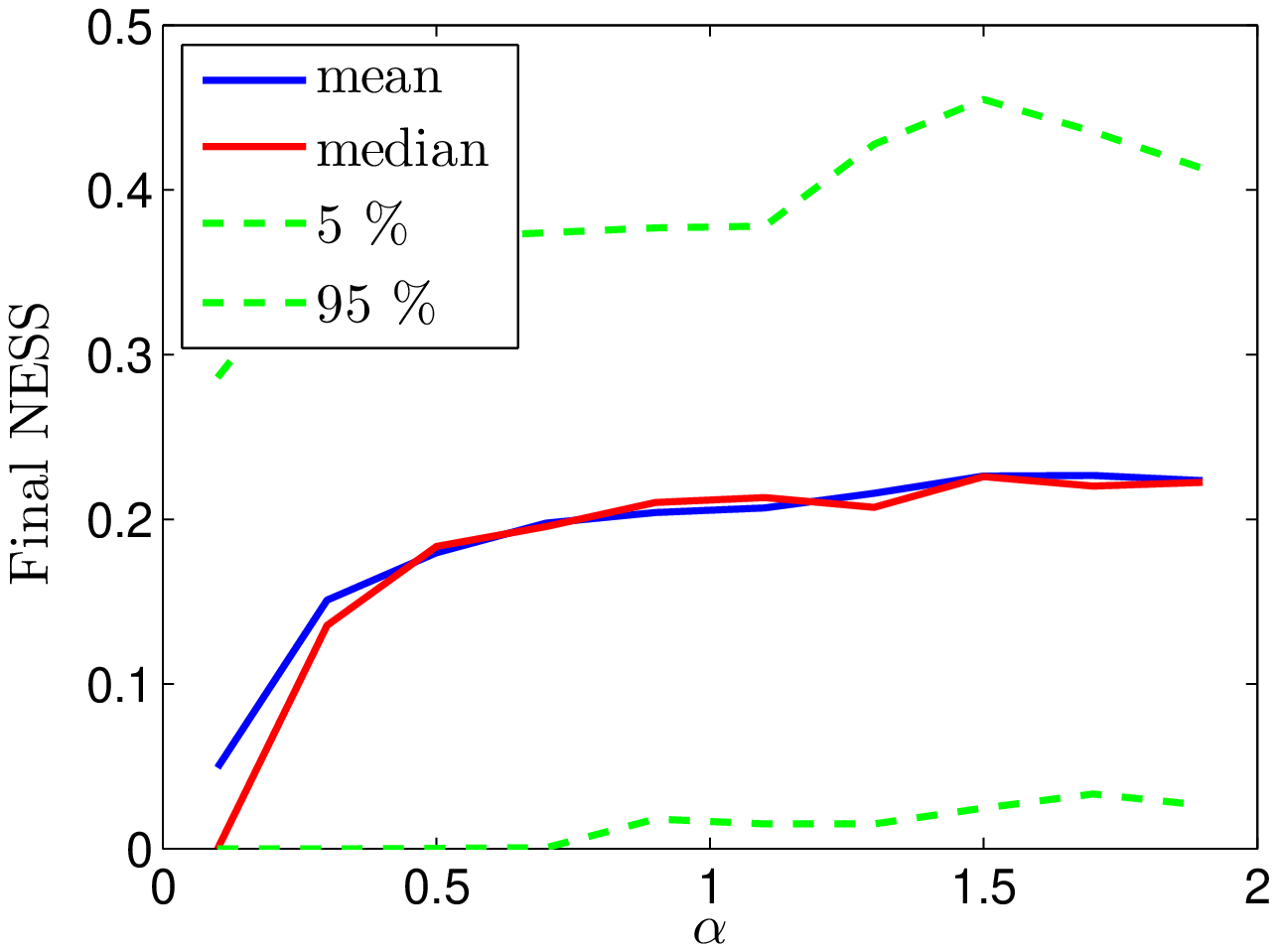}
\includegraphics[width=0.49\textwidth]{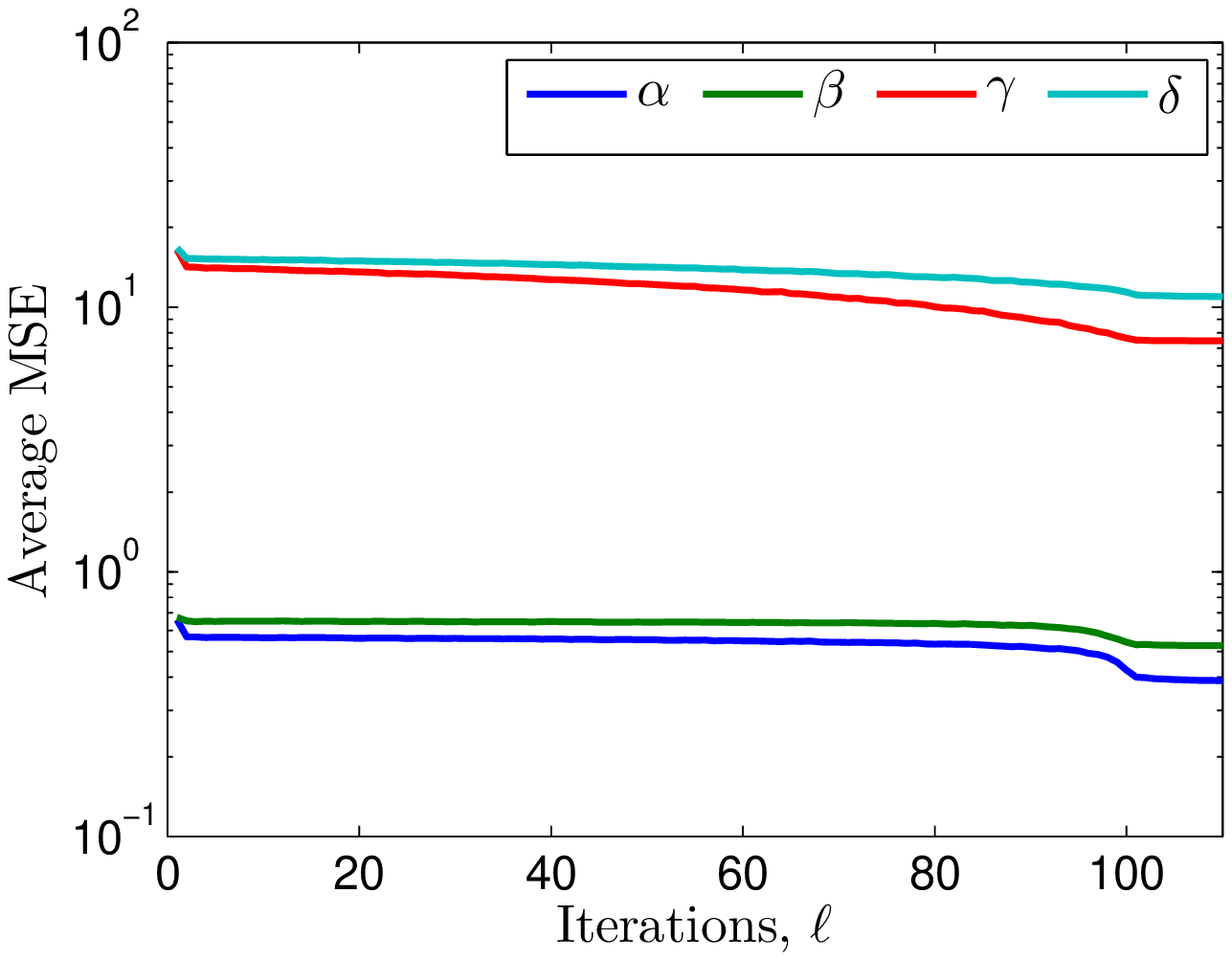}
\caption{\textit{Left}: Final NESS statistics versus the true value of $\alpha$ obtained by the PMC-ABC algorithm with the prior distribution $p_1(\bftheta)$. \textit{Right}: Evolution along the iterations of the average MSE obtained by the PMC-ABC method with prior $p_1(\bftheta)$.}
\label{fig_4}
\end{figure*}

In Figure \ref{fig_4} we present the results obtained by the PMC-ABC method under the prior distribution $p_1(\bftheta)$. The \textit{left} plot shows the statistics (mean, median, 5\% and 95\% quantiles) of the final NESS at the final iteration of the PMC-ABC algorithm. In this case, the NESS is computed in the same manner as for the NPMC method. It can be observed that, particularly for low $\alpha$, the final NESS takes very low values, around 0.2 on average in the best case. In the \textit{right} plot, the evolution of the average MSE is represented versus the iteration index $\ell$. Only a slight improvement in terms of MSE can be observed along the iterations. If we further reduce the threshold $\epsilon_\ell$ in order to improve the likelihood approximation, the computational time shoots up and the NESS values drop, leading to numerical instabilities. The results obtained with the broader prior $p_2(\bftheta)$ are extremely poor and have been omitted.

\subsection{Comparison of the Bayesian methods}
\label{comp_Bayesian}

\begin{figure*}[ht]
\centering
\includegraphics[width=0.49\textwidth]{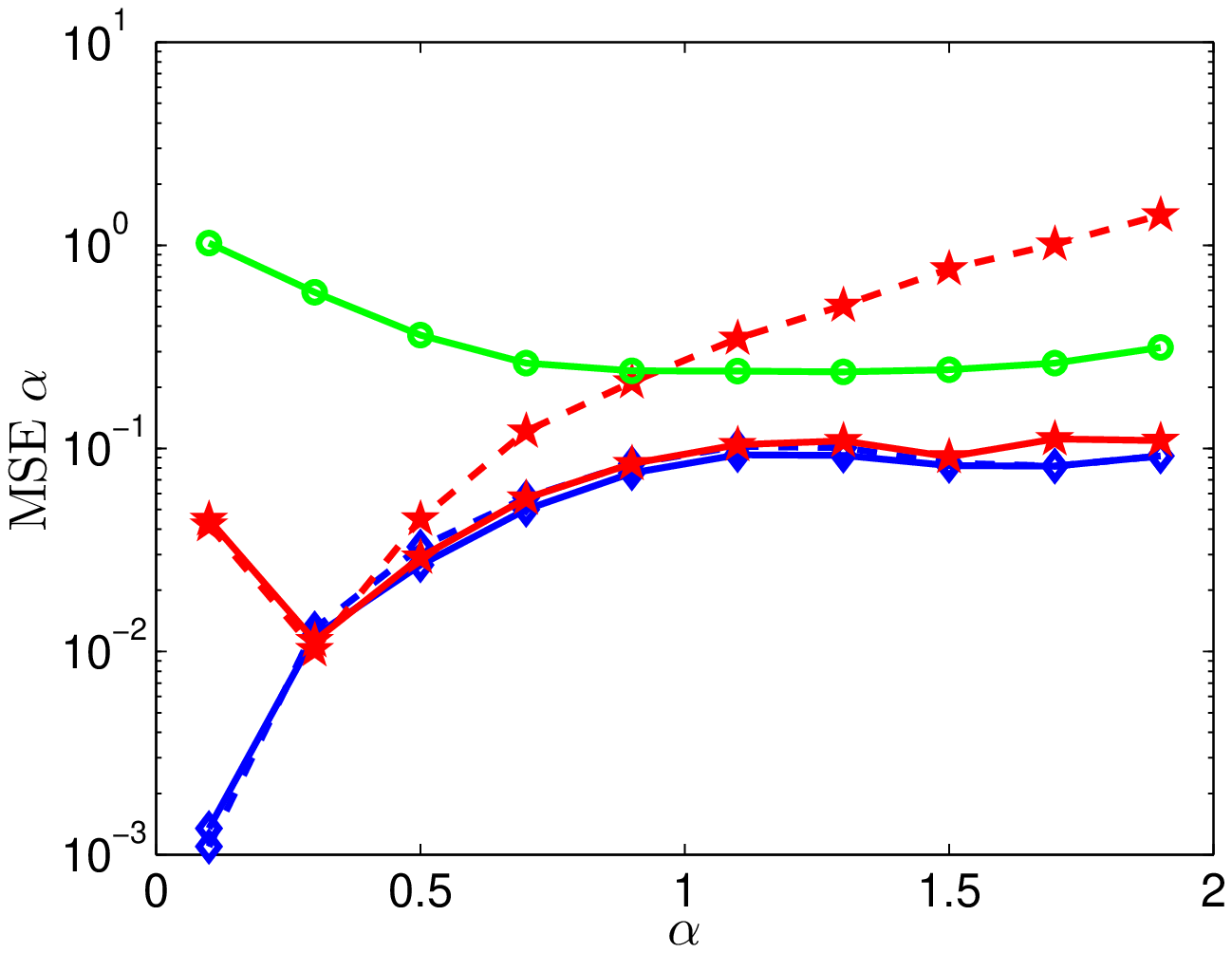}
\includegraphics[width=0.49\textwidth]{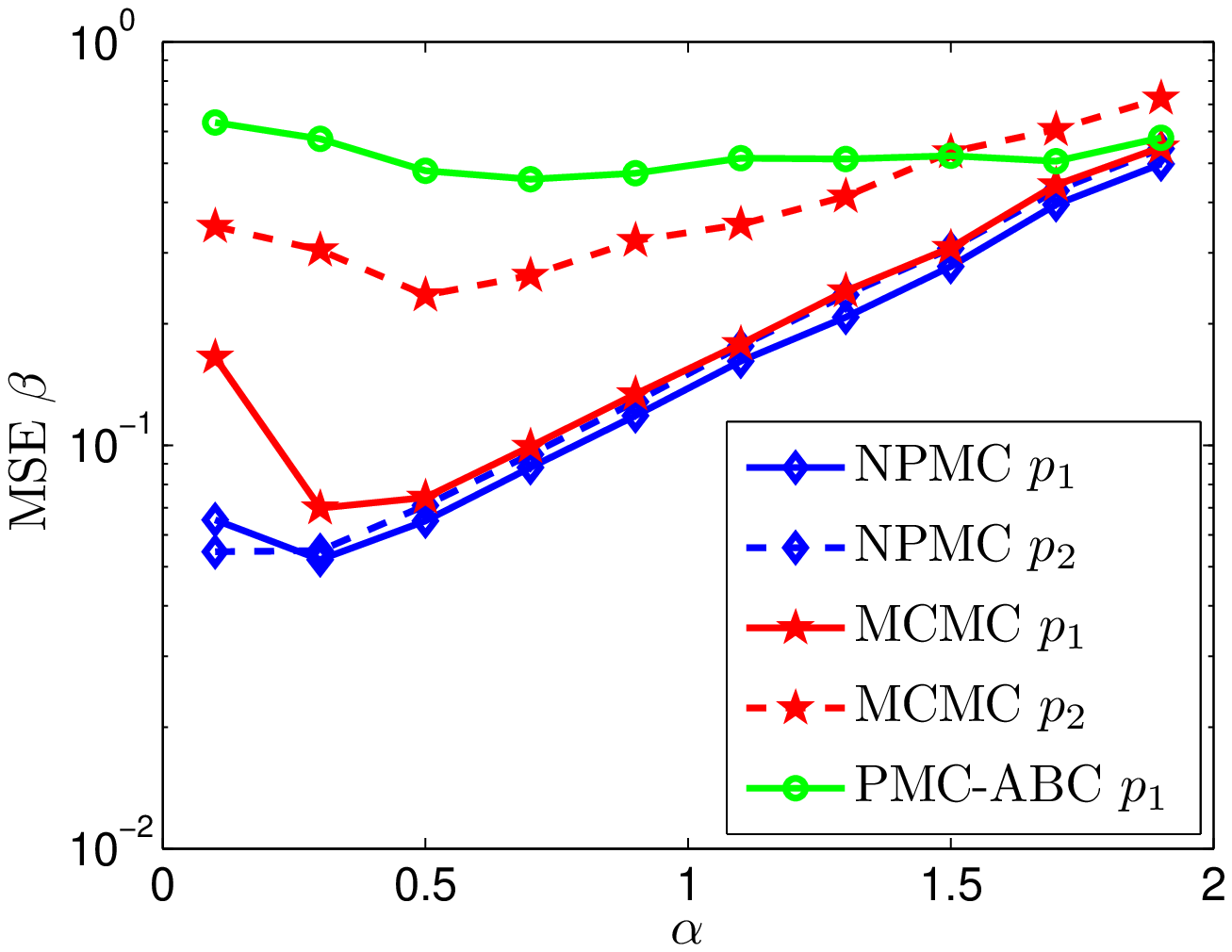}
\includegraphics[width=0.49\textwidth]{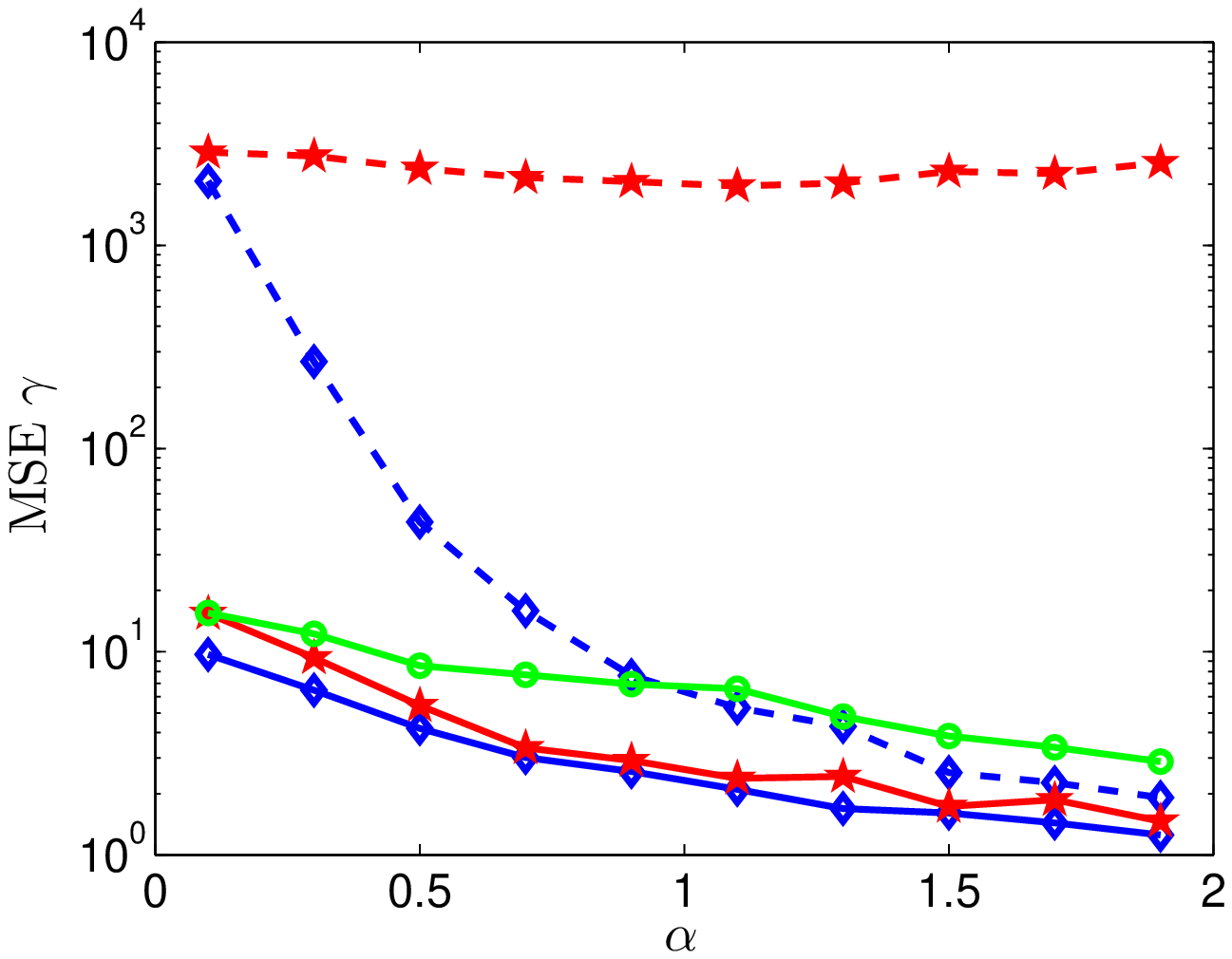}
\includegraphics[width=0.49\textwidth]{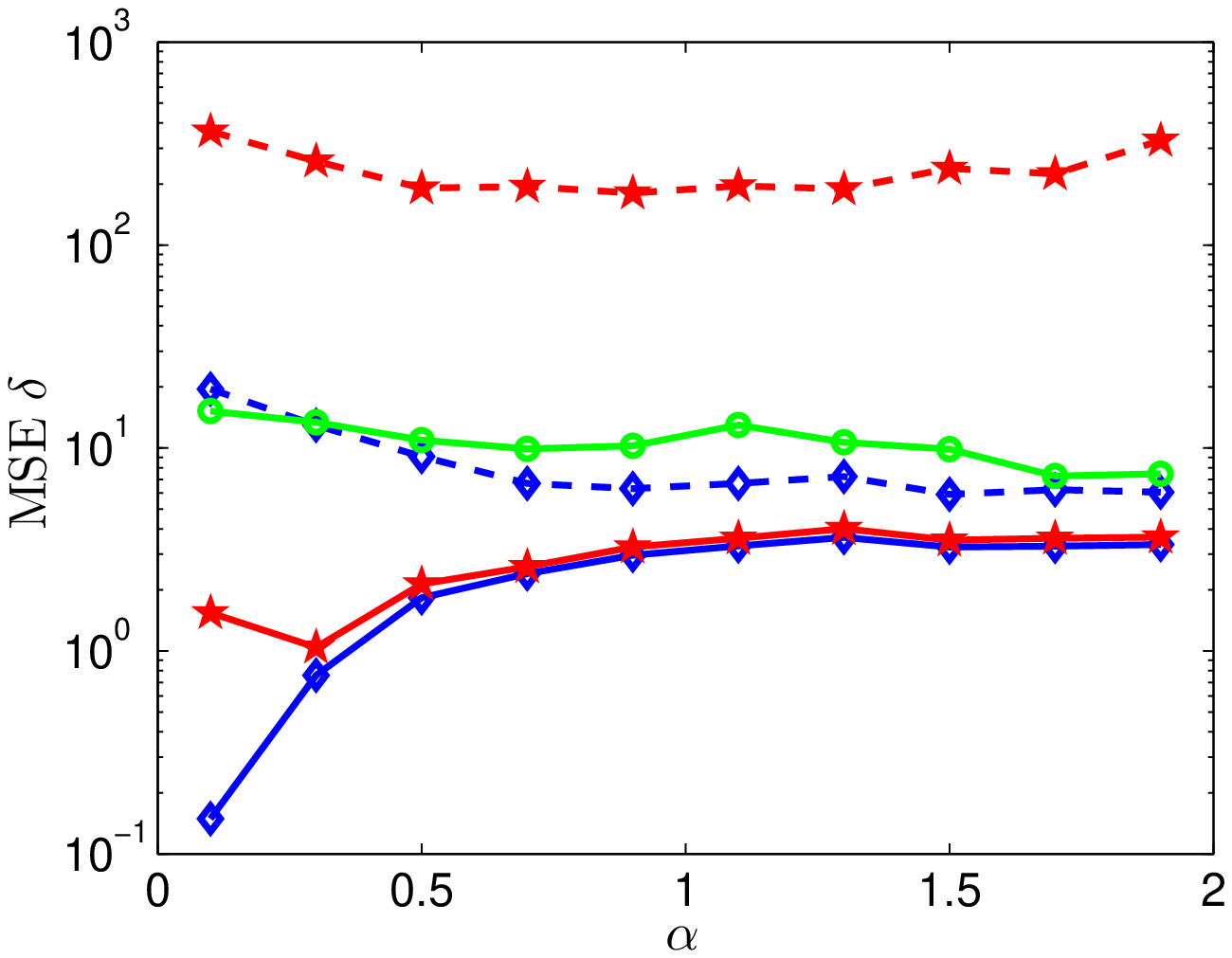}
\caption{Average final MSE of each parameter versus the true value of $\alpha$, obtained by the NPMC and MH methods, with the $p_1(\bftheta)$ and $p_2(\bftheta)$ prior distributions and 5000 simulations, and by the PMC-ABC method with $p_1(\bftheta)$ and 2500 simulations. The curves have been obtained by averaging the final MSE obtained in each simulation run in intervals of $\alpha$ of length 0.2.}
\label{fig_5}
\end{figure*}

In Figure \ref{fig_5}, the final average MSE of each parameter is represented versus the true value of $\alpha$, as obtained by the NPMC and the MH methods with both prior choices $p_1(\bftheta)$ and $p_2(\bftheta)$, and by the PMC-ABC method only with prior $p_1(\bftheta)$. The MSE has been computed from the final sample, taking into account both the bias and the variance of the estimates, since the full posterior approximation allows to do so. It can be observed that both the NPMC and the MH techniques perform similarly with the prior distribution $p_1(\bftheta)$, except for $\alpha < 0.2$, where the NPMC attains better results. However, when the broader prior $p_2$ is considered, the MH algorithm yields highly inaccurate results due to the inefficiency of the Markov chains to explore the broader space of $\bftheta$ (which leads to low acceptance rates and a high correlation among samples). This leads to a minor MSE reduction w.r.t. the prior distribution, especially for $\gamma$ and $\delta$. Much longer chains would be required to obtain reasonable results with this prior distribution. On the contrary, the NPMC method obtained similar MSE values in the estimation of $\alpha$ and $\beta$ with both prior choices, for any value of $\alpha$. The $\gamma$ and $\delta$ parameters present significantly worse performance with the broader prior $p_2$, especially for low values of $\alpha$. This reveals that with the low amount of observations considered in this setting ($T=30$), the $\gamma$ (and, to a lesser extent, $\delta$) parameter cannot be identified when the distribution of interest presents very heavy tails. The selection of an informative prior for $\gamma$ and $\delta$ leads to more efficient and robust algorithms, since it avoids the overestimation of these parameters, and allows to reduce the number of required samples. The likelihood-free method performs poorly compared to any of the other Bayesian techniques, for all $\alpha$. In view of these results, the NPMC algorithm appears to clearly outperform the other Bayesian methods. 

It is of particular interest to understand the behavior of the addressed Bayesian algorithms when $\alpha$ takes low values, in particular when $\alpha < 0.5$. When $\alpha$ is close to zero, it is well known that the resulting $\alpha$-stable density presents an extremely sharp mode and heavy tails. 
Very large variations of the likelihood function prevent standard Bayesian methods from performing properly, especially when the prior pdf is broad with respect to the likelihood. For instance, it often results in a very high rejection rate for MCMC algorithms, which require very long chains to provide reasonable estimates. Likelihood-free methods are particularly inefficient and turn out to be of no practical use in this case. Standard IS techniques also suffer from this problem, and require very large sample sizes due to the weight degeneracy. However, the proposed NPMC algorithm specifically addresses the degeneracy problem and is especially suited for problems where the mode of the target pdf is very sharp. Large weight variations are handled by the clipping procedure, ensuring a robust performance in the difficult case when $\alpha$ is very low.

\subsection{Comparison of the NPMC algorithm with non-Bayesian methods}
\label{comp_noBayesian}

In this section we provide a comparison of the performance of the NPMC method with some of the main non-Bayesian methods proposed in the literature.
Specifically, we consider the classical quantile method proposed in \cite{Mcculloch1986} (QT1), the modified quantile method in \cite{Nolan2013} (QT2), the ECF-based method of \cite{Kogon1998} (ECF), the ML estimation method of \cite{Nolan2001} (MLE) and the log-absolute moments method proposed in \cite{Nikias1995} (LAM). 
All of these methods are implemented in the toolbox STABLE for different platforms, and provide point estimates $\hat{\theta}_k$ of the $\alpha$-stable parameters from a set of observed data. We have performed $10^5$ independent simulations of each of these methods and computed the empirical MSE from the point estimates of each parameter $\theta_k$, $k=1,\ldots,4$, as $MSE_k = (\hat{\theta}_{k} - \theta_{k})^2$. For the NPMC method, we have obtained the point estimate as the approximate posterior mean obtained at the last iteration $L=10$, i.e., $\hat{\theta}_k = \mu_{L,k}$ (ignoring the variance of the estimator), and thus the curves slightly differ from those shown in Figure \ref{fig_5}. In the case of the NPMC technique we have considered 5000 simulation runs with the prior distribution $p_1(\bftheta)$. The simulation setup regarding the generation of the observed data fits the one described in Section \ref{Sims_NPMC}.

\begin{figure*}[ht]
\centering
\includegraphics[width=0.49\textwidth]{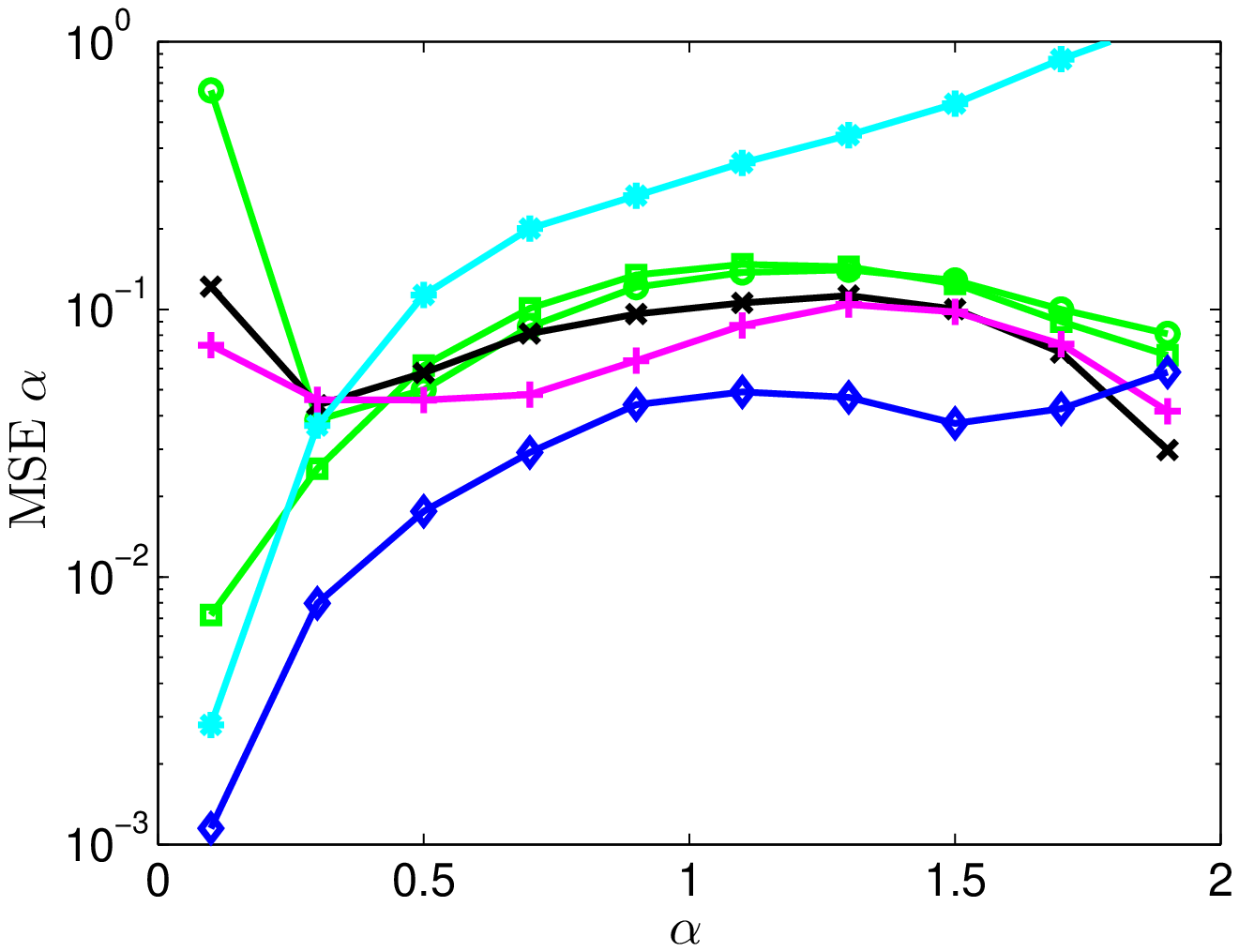}
\includegraphics[width=0.49\textwidth]{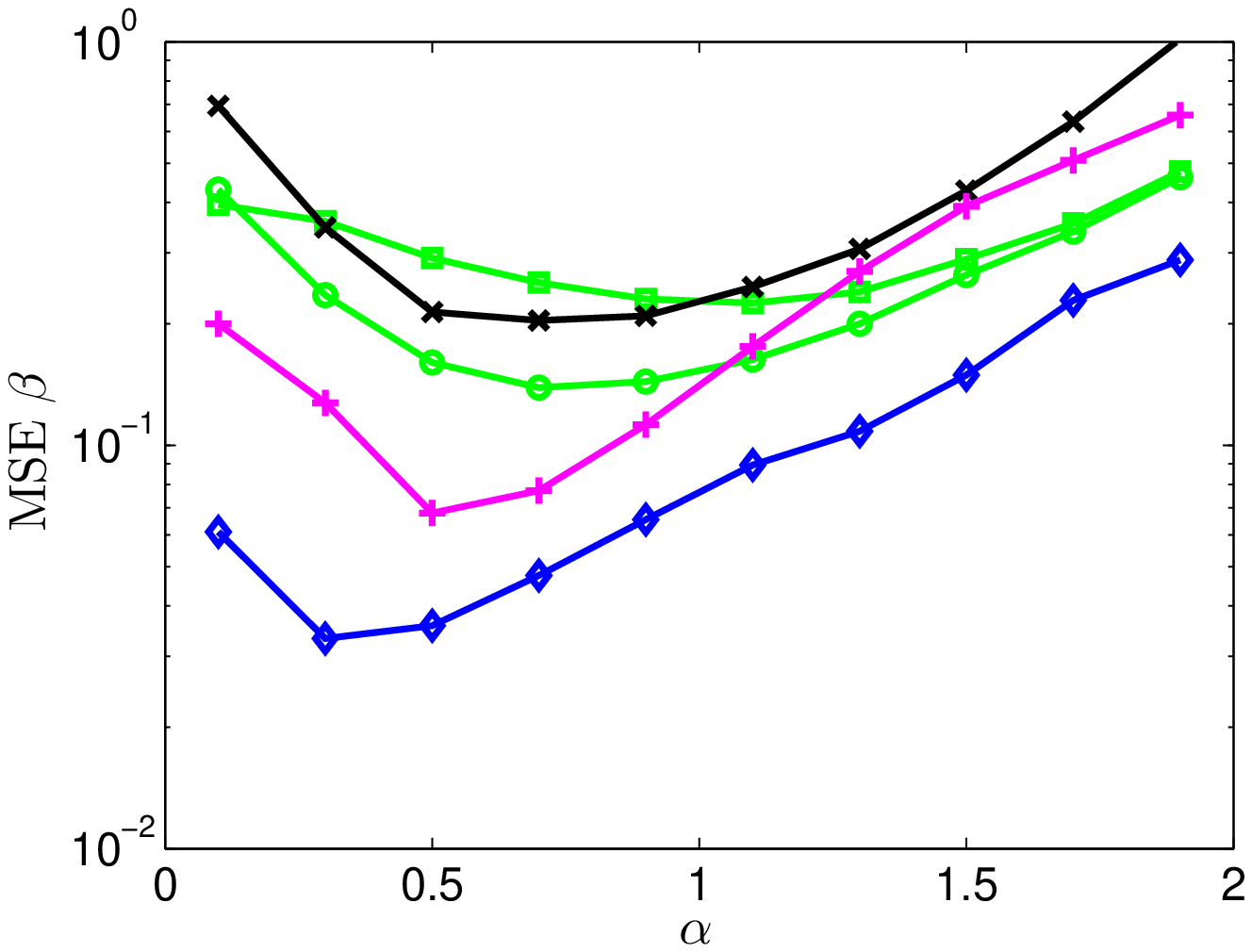}
\includegraphics[width=0.49\textwidth]{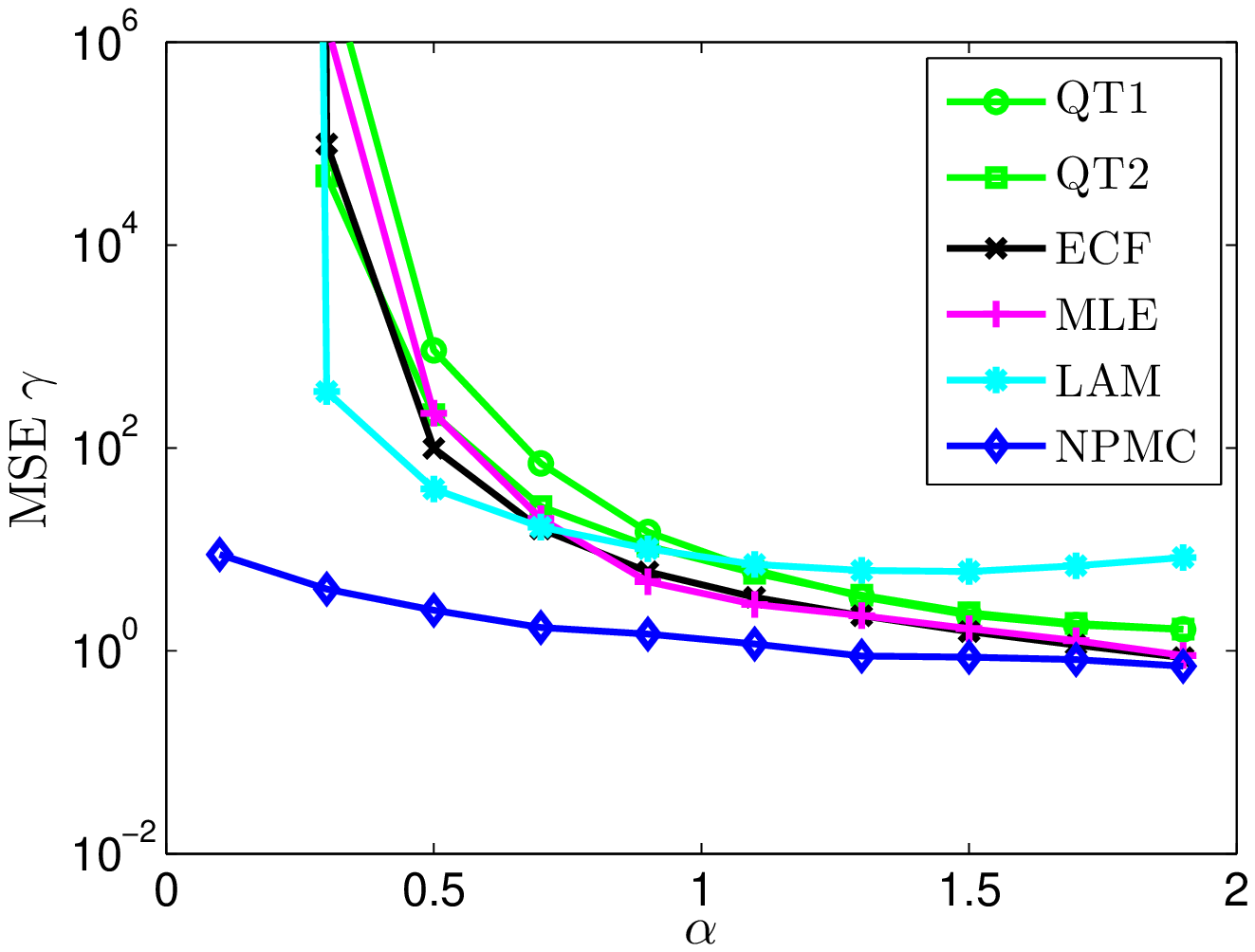}
\includegraphics[width=0.49\textwidth]{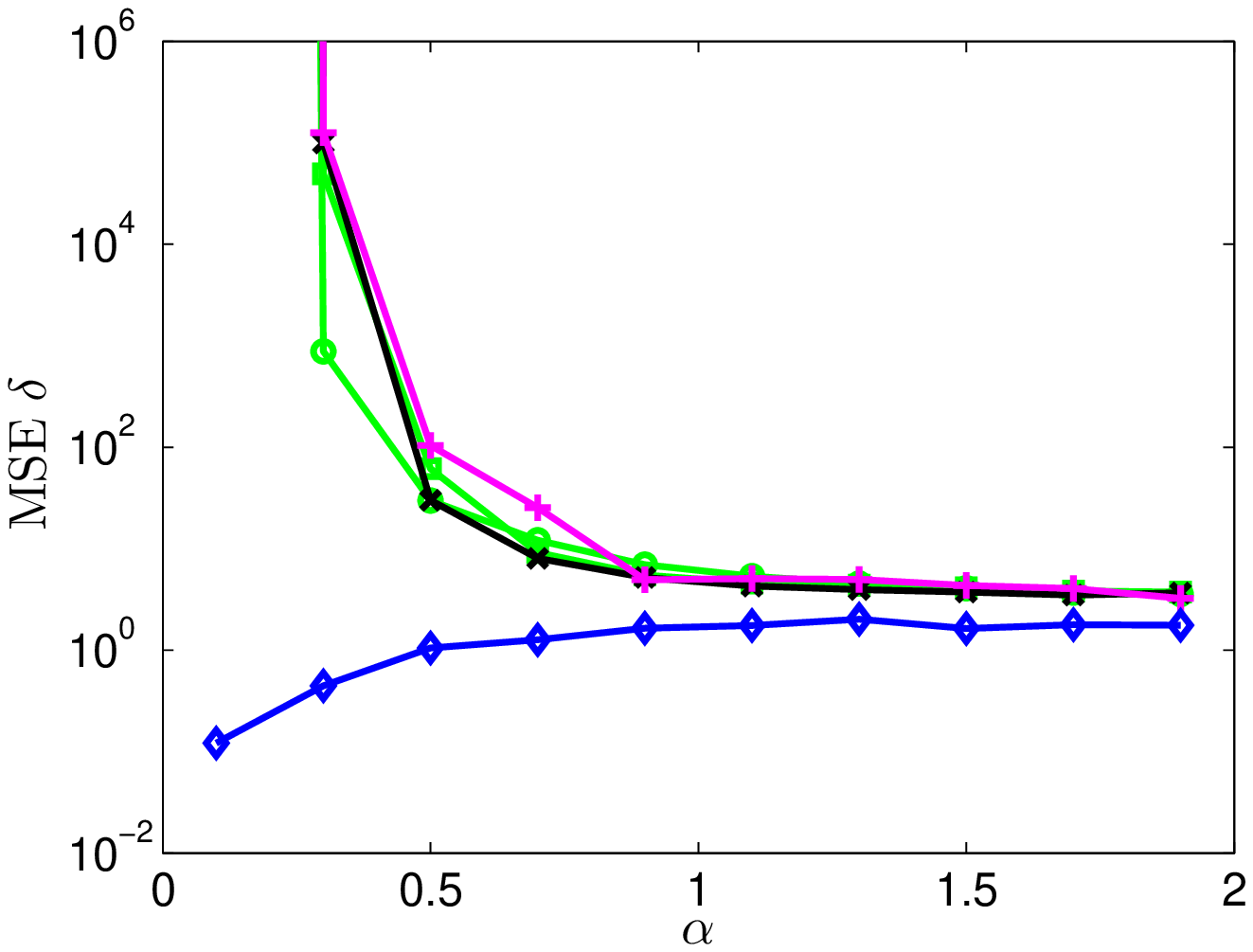}
\caption{Empirical MSE of each parameter averaged over $10^5$ independent simulation runs versus the true value of $\alpha$, obtained by the QT1, QT2, ECF, MLE, LAM and NPMC methods. The curves have been obtained by averaging the empirical MSE values obtained in each simulation run within intervals of $\alpha$ of length 0.2. The curves of the NPMC correspond to the narrow prior $p_1$.}
\label{fig_6}
\end{figure*}

In Figure \ref{fig_6}, the final MSE obtained by the various methods for each parameter and averaged over $10^5$ independent simulation runs is represented versus the true value of $\alpha$. The LAM technique only provides estimates for $\alpha$ and $\gamma$. Regarding the estimation of the $\alpha$ parameter, the QT1, ECF and MLE methods are unable to estimate values of $\alpha < 0.4$. On the contrary, the NPMC, the QT2 and the LAM methods succeed to estimate low values of $\alpha$. The NPMC method outperforms the other methods for all values of $\alpha$, except for $\alpha \approx 2$, which corresponds to a Gaussian distribution. We have observed that the posterior mean can yield an underestimation of $\alpha = 2$, and that the posterior mode can be more appropriate in this case. For the estimation of $\beta$ the NPMC method also provides the best results, followed by the MLE and the QT1 methods. Given the low amount of observations, all of the methods fail to accurately estimate the true values of $\gamma$ and $\delta$ for $\alpha < 0.5$, yielding the NPMC method the best (yet modest) results. 

\subsection{Remarks}

The results provided in Sections \ref{comp_Bayesian} and \ref{comp_noBayesian} show that the NPMC method outperforms all the other methods we have studied (both Bayesian and frequentist) in terms of MSE for all $\alpha$. Additionally, the NPMC method is more robust to numerical issues, occurring mainly with low values of $\alpha$. In Table \ref{tabla_fallos} the failure rate of each method is shown, together with the corresponding execution times. The failure rate is defined as the percentage of simulation runs that end with a numerical error or warning indicating that the provided results are inaccurate. The QT1, ECF, NPMC and MH methods are very robust to the $\alpha$ parameter value and only fail in around 0.35\% of the simulations, when $\alpha < 0.01$. However, the MH algorithm performs poorly with the broader prior $p_2(\bftheta)$, yielding a high failure rate. The MLE method provides an error rate over 20\% because the likelihood approximation routine implemented in STABLE does not work for $\alpha < 0.4$. The LAM method fails in 8\% of the simulations, probably due to the low amount of observations considered, especially for heavy-tailed distributions. Finally, for the PMC-ABC method the failure rate is expressed in terms of the number of simulations that did not reach iteration $\ell=50$ by the time limit of 15 minutes.

\begin{table}[ht] \centering
\caption{Failure rate and execution time of each algorithm.}
\label{tabla_fallos}
\begin{tabular}{c|ccccc|ccc}
  \hline
   & QT1 & QT2 & ECF & MLE & LAM & NPMC & MH & PMC-ABC \\
  \hline
  Failure rate (\%)& 0.37 & 5.05 & 0.37 & 21.1 & 7.89 & 0.35 &  0.5 & 27 \\
	Execution time & $<$ 1 sec & $<$ 1 sec & $<$ 1 sec & $<$ 1 sec & $<$ 1 sec & 5 min & 5 min & 15 min \\
   \hline
\end{tabular}
\end{table}

Regarding the execution times, Bayesian methods are significantly slower than the classical frequentist techniques. The NPMC and MH methods have similar computational complexity, while the ABC method is much slower. We have used the R version of Nolan's STABLE 4.0 \cite{WebNolan} to run the non-Bayesian techniques included in the comparison. All Bayesian methods have been implemented in Matlab R2007b on a 3-GHz Intel Core 2 Duo CPU E8400, with 2 GB of RAM. Contrary to the MH algorithm, in the case of NPMC and PMC-ABC, the processing of each sample in a given iteration can be easily parallelized to reduce the running time. Any of the Bayesian methods can incorporate additional information in the prior pdf based on point estimates such as MLE or QT2, in order to reduce the number of required iterations. Also note that the NPMC algorithm only requires around 5 or 6 iterations for convergence on average, which can reduce the execution time by one half with a very slight loss of performance.

The execution time of both NPMC and MCMC algorithms is almost exclusively dedicated to evaluating the likelihood function (99\% of the simulation time). The computational effort dedicated to generating samples, computing the IWs or the acceptance rates is negligible compared to the likelihood evaluation.  If a more accurate or faster pdf evaluation becomes available, the NPMC (and MCMC) algorithm can directly benefit from this performance improvement.

It has to be noted that some of the frequentist methods, especially ECF and MLE, provide reasonable estimates of all 4 parameters with little computational complexity whenever $\alpha > 0.3$. However, the NPMC algorithm yields more accurate estimates in general, and performs significantly better for very low $\alpha$, at the expense of an increase in the execution time.

In comparison with other Bayesian methods, the NPMC algorithm has clear advantages in terms of simplicity, estimation error and execution time. The NPMC method is straightforward to implement, and it only requires a coarse selection of the parameters $L$, $M$ and $M_T$. We propose to use $M_T \approx \sqrt{M}$, according to the theoretical convergence results given in Section \ref{Analysis} and \cite{Koblents2013a}. The convergence of the NPMC may be easily assessed in practice by observing the evolution of the NESS along the iterations, and stopping the adaptation when it reaches a steady value.
Additionally, the NPMC method scales better as the complexity of the problem increases (a broader prior or a larger number of observations).

The advantages of the proposed algorithm are most apparent when estimating the parameters of extremely heavy-tailed distributions (very low $\alpha$) from small data sets. When the number of observations is larger, all Bayesian methods become computationally more demanding. For example, with $T=300$ all the methods yield more accurate estimates, but the running time of the NPMC and MH algorithms scales up to 13 minutes. However, the MLE estimates barely improve for low values of $\alpha$, even with a larger amount of observations, due to the coarse likelihood approximation implemented in the STABLE toolbox.


\section{Simulations with real fish displacement data}
\label{Sims_peces}

In this section we present the numerical results obtained with a set of real data describing longitudinal (i.e., upstream or downstream) movements of fish in a stream. This data set was first described in \cite{Belanger2001}.

\subsection{Data description}

The data analyzed here derive from measurements of daytime (09:00 - 17:00 hours) position made on $N=21$ individuals of a single fish species, the brook trout (\textit{Salvelinus fontinalis}). Fish position, determined by reference to fixed marks along the stream edge, was tracked by radiotelemetry daily between 29 July and 5 September 1998 in Ganelon Creek, Canada. The available set of observations $\textbf{y}$ corresponds to the univariate daily longitudinal displacement of individual fish, measured in meters. The $t$-th displacement of the $n$-th fish, $y_{n,t}$, is defined as the position increment in one dimension between two consecutive measurements. 

\begin{figure*}[ht]
\hspace{-0.4cm}
\includegraphics[width=0.25\textwidth]{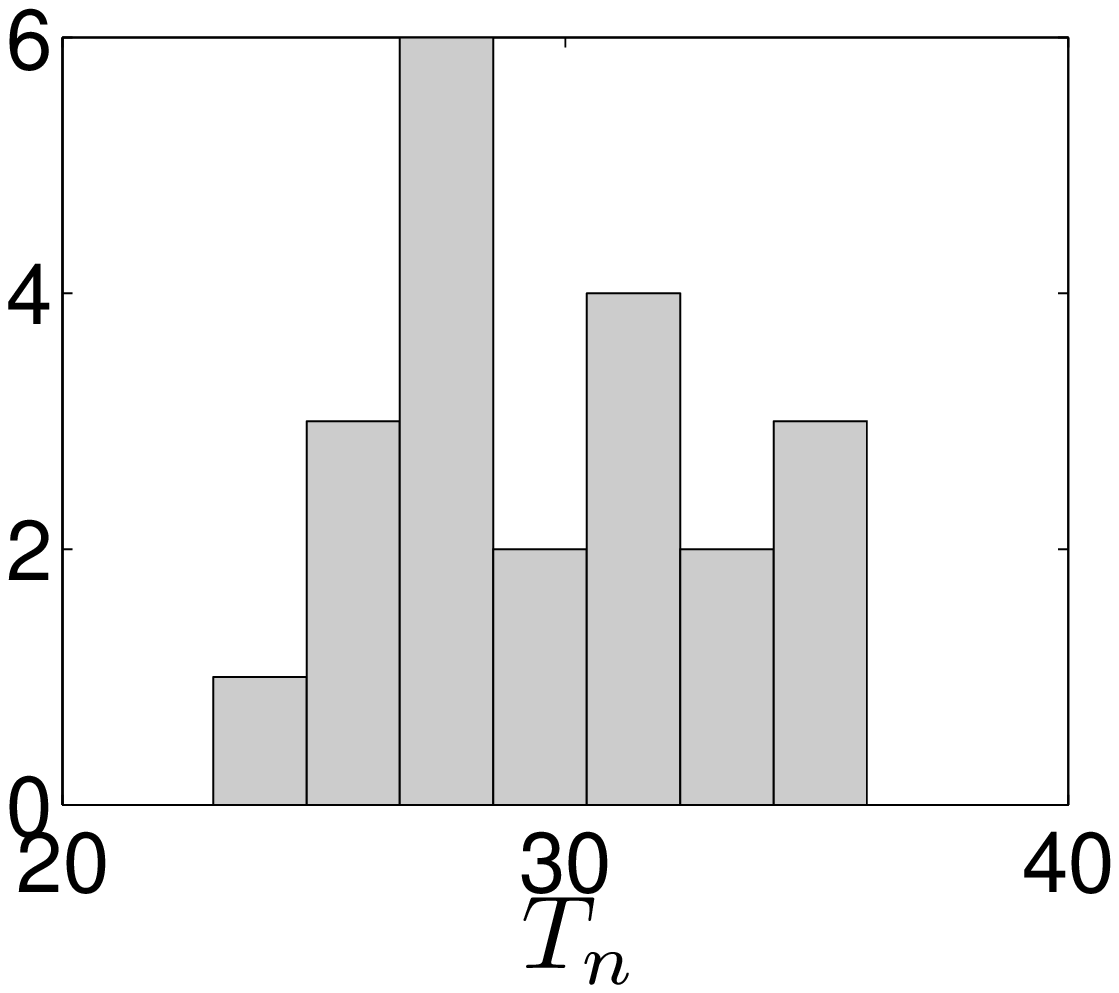}
\includegraphics[width=0.25\textwidth]{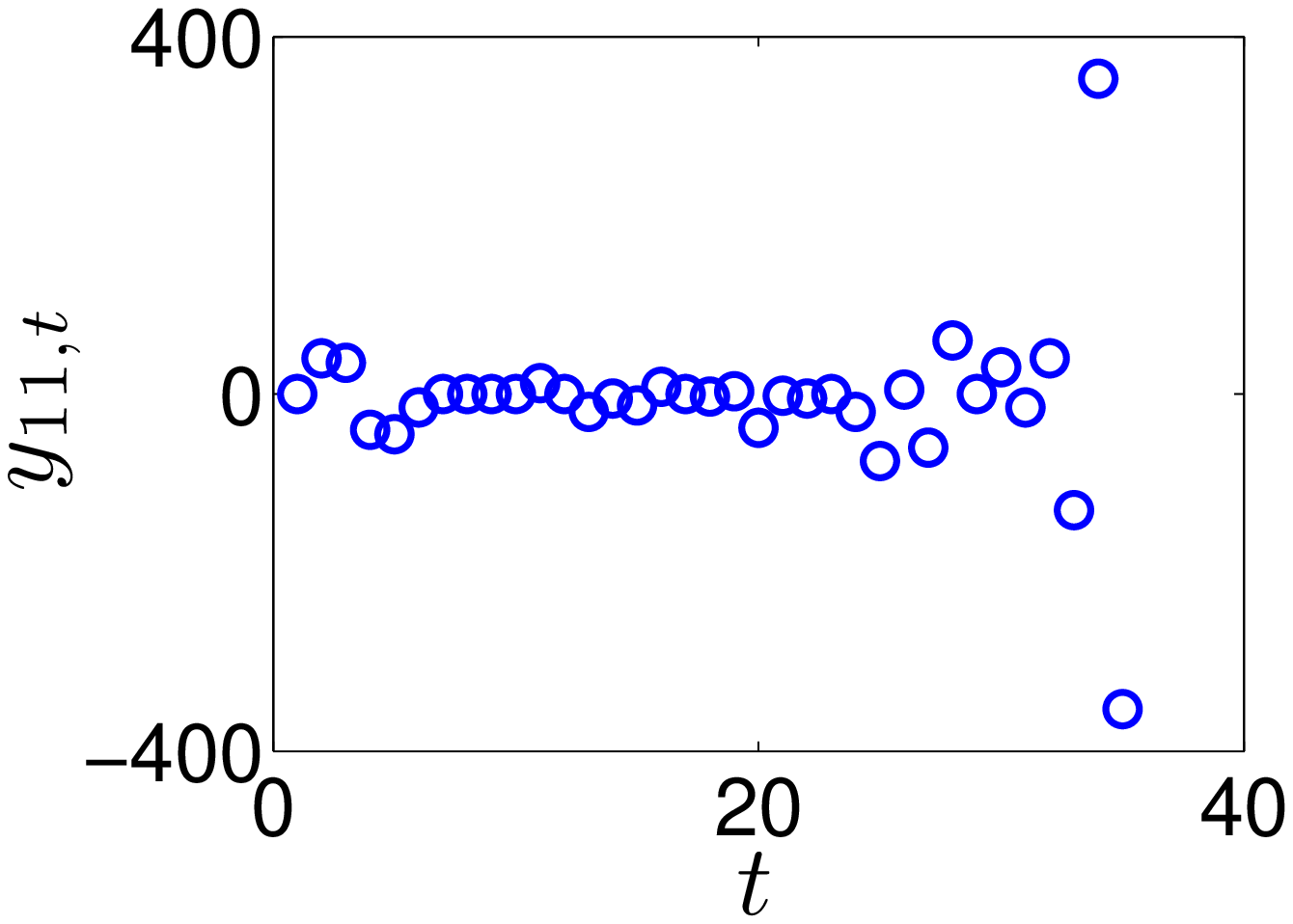}
\includegraphics[width=0.25\textwidth]{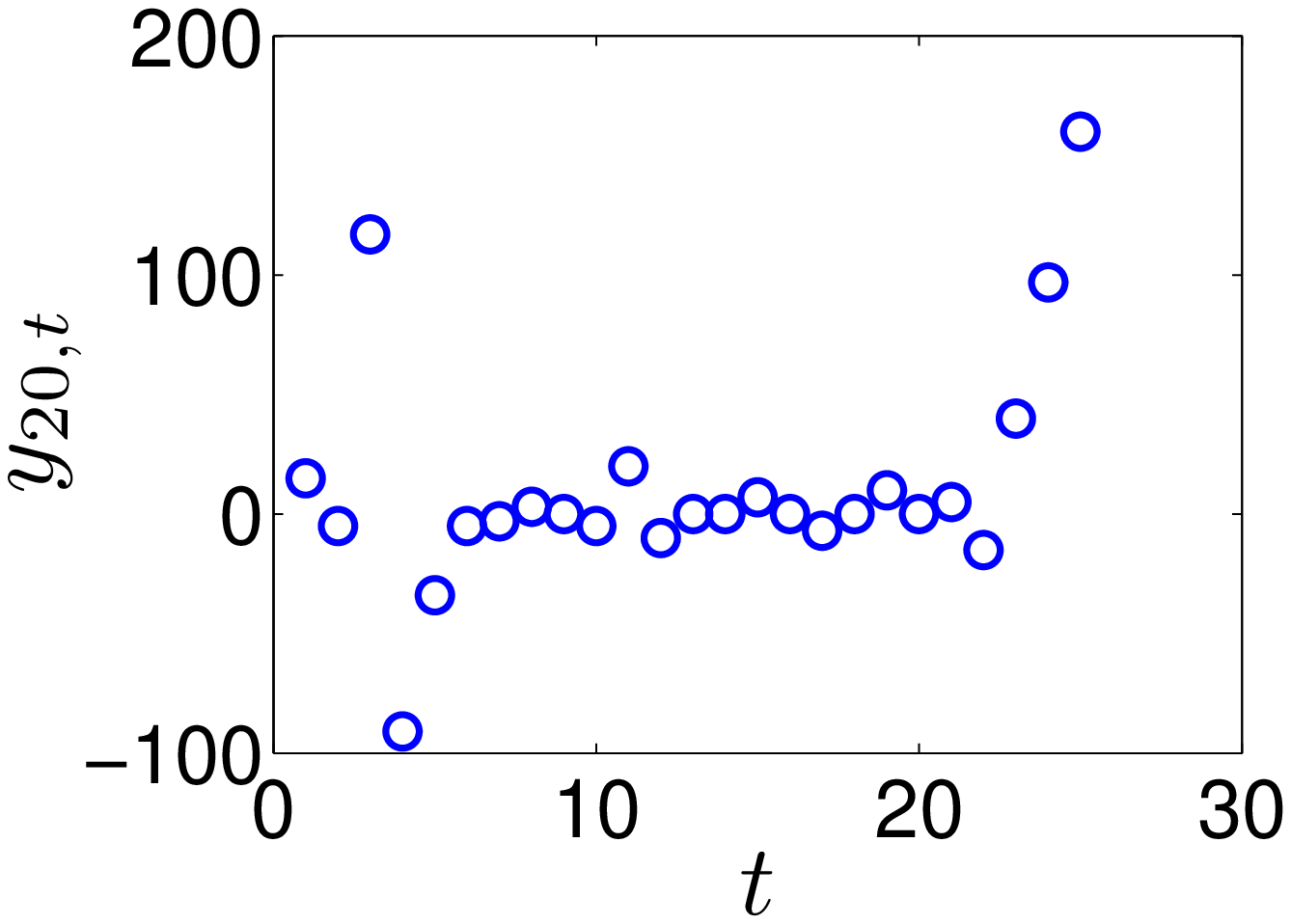}
\includegraphics[width=0.25\textwidth]{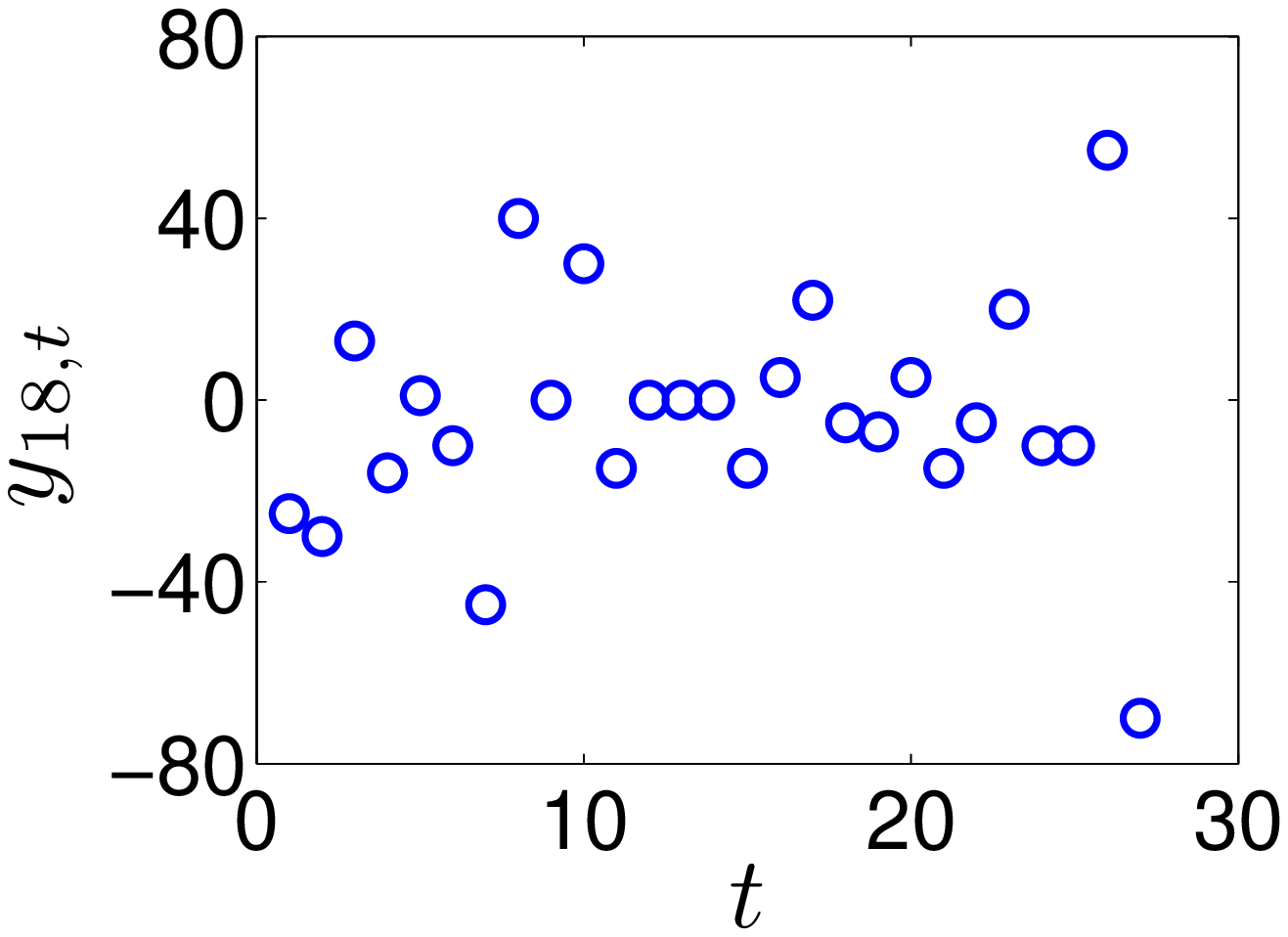}
\caption{First plot from the left: histogram of the number of observations $T_n$ available for each fish. Second to fourth plots from the left: real measurements of fish displacement $y_{n,t}$, $t=1, \ldots,T_n$, of three selected individuals $n=11, 20, 18$ with different characteristic behaviors.} \label{fig_7}
\end{figure*}

The number of observations $T_n$ associated with each individual is very small and its histogram is shown in Figure \ref{fig_7} (\textit{left}), together with the available observations $y_{n,t}$ of three selected individuals, $n=11,20,18$, at each time instant $t$. These three cases describe the typical behaviors present in the whole available data set. Although displacements were quantified under similar environmental conditions, individuals were analyzed separately because heterogeneity in individual behavior generally is of ecological interest. For example, the estimates of $\beta$ could be used to assess interindividual differences in the propensity to exhibit directional movements, such as upstream or downstream migration.

Visual inspection of the available data reveals that it has no gaps in its support and presents unimodality, heavy-tails and asymmetry, and cannot be properly modeled by a Gaussian distribution \cite{Nolan1999fitting}. Thus, we assume that these samples are independent and follow an $\alpha$-stable distribution $y_{n,t} \sim \mathcal{S}(y; \alpha_n, \beta_n, \gamma_n, \delta_n)$. The individual $n=11$ (second plot from the left in Figure \ref{fig_7}) presents a heavy-tailed and rather symmetric distribution, probably with a low value of $\alpha$ and $\beta$. Fish $n=20$ (third plot from the left in Figure \ref{fig_7}) presents lighter tails and more asymmetry than the previous case. Finally, $n=18$ (first plot from the right in Figure \ref{fig_7}) corresponds to a light-tailed and apparently symmetric distribution, similar to a Gaussian population.

\subsection{Numerical results}

\begin{figure*}[ht]
\begin{centering}
\label{fig_8}
\includegraphics[width=0.45\textwidth]{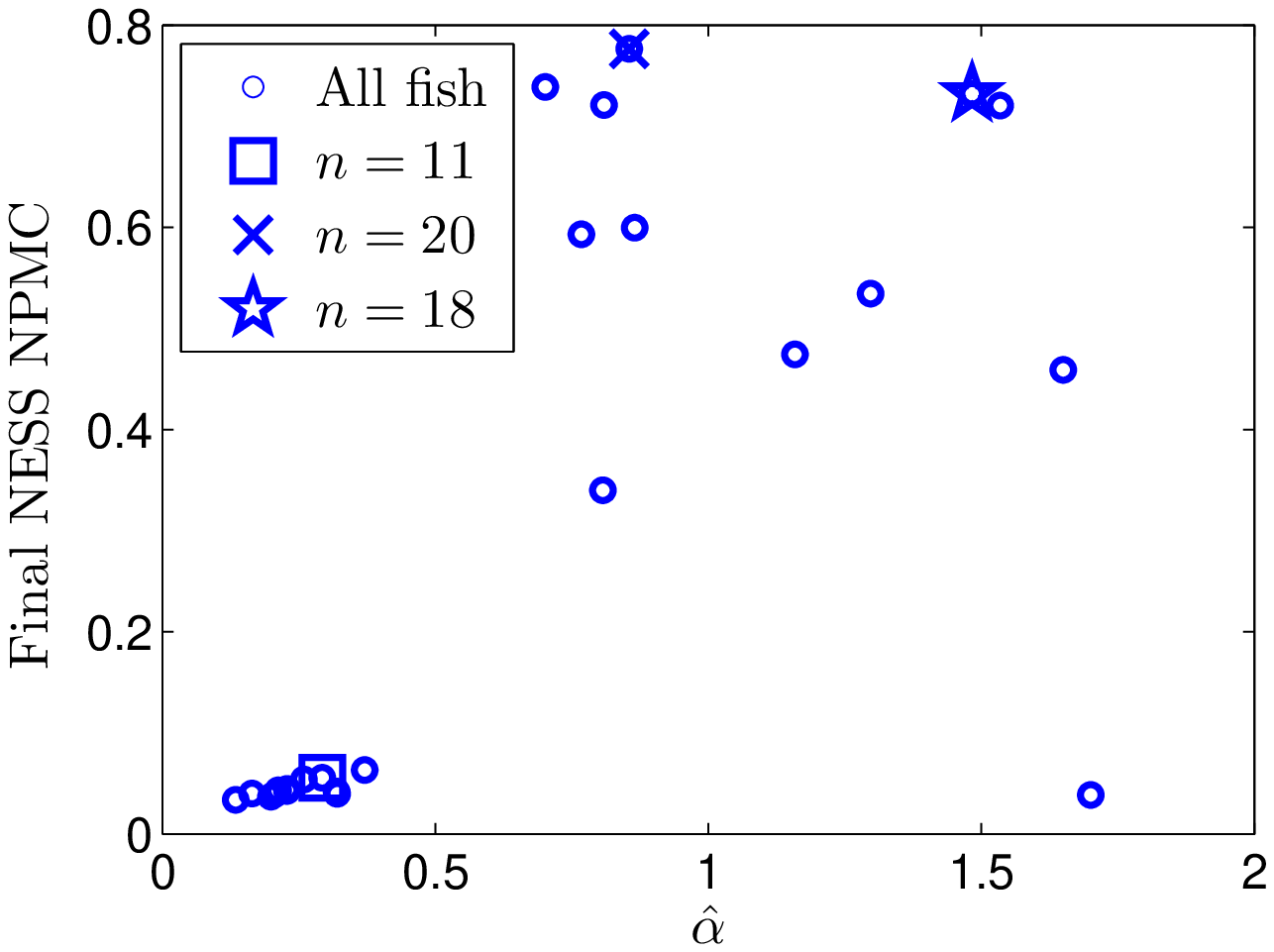}
\includegraphics[width=0.45\textwidth]{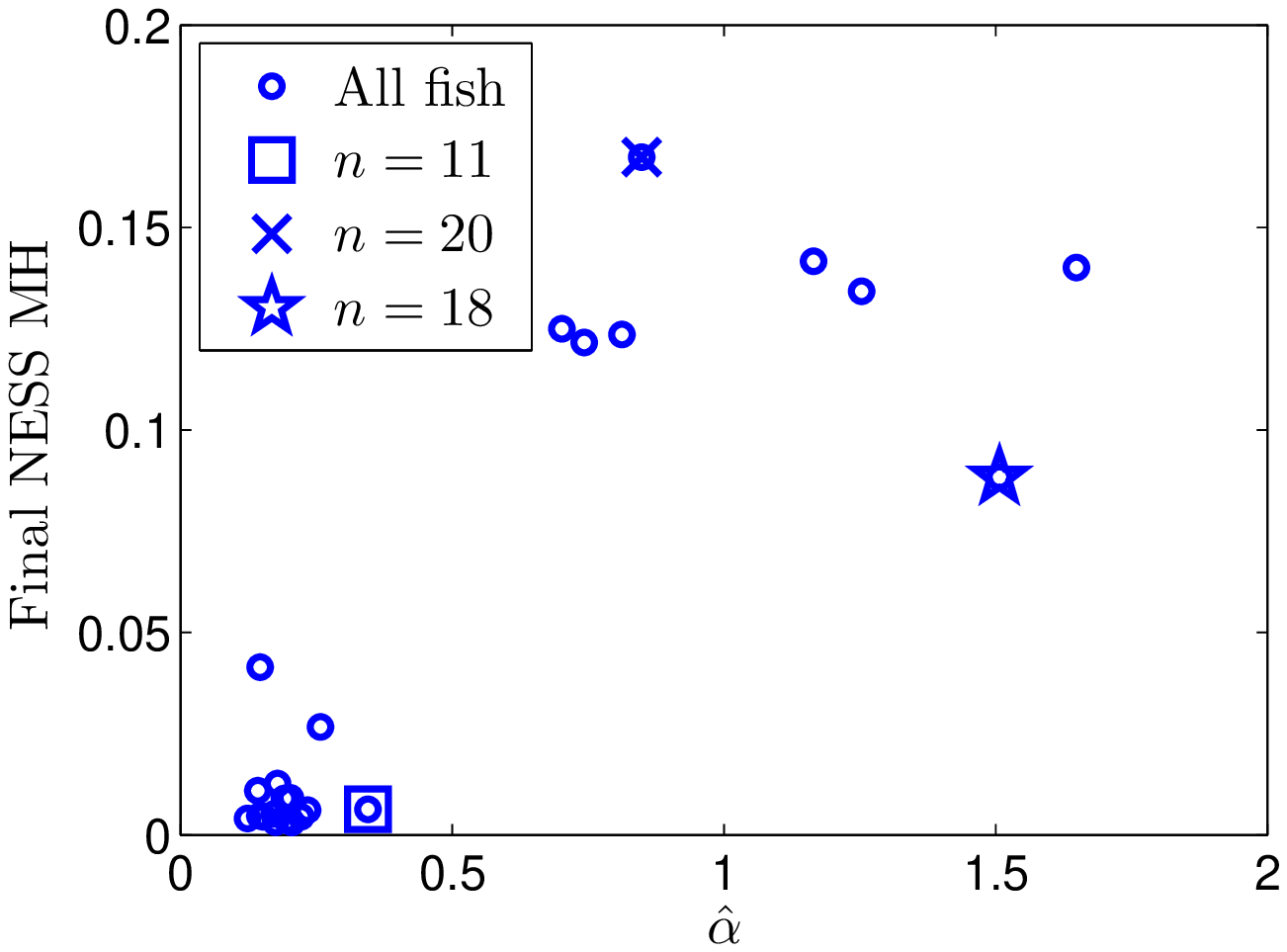}
\vspace{-0.2cm}\caption{Final NESS obtained in each simulation by the NPMC (\textit{left}) and the MH (\textit{right}) algorithms, versus the corresponding estimates of $\alpha$. Note that the NESS is computed differently in both cases.}
\end{centering}

\vspace{0.3cm}

\hspace{-2cm}
\includegraphics[width=1.2\textwidth]{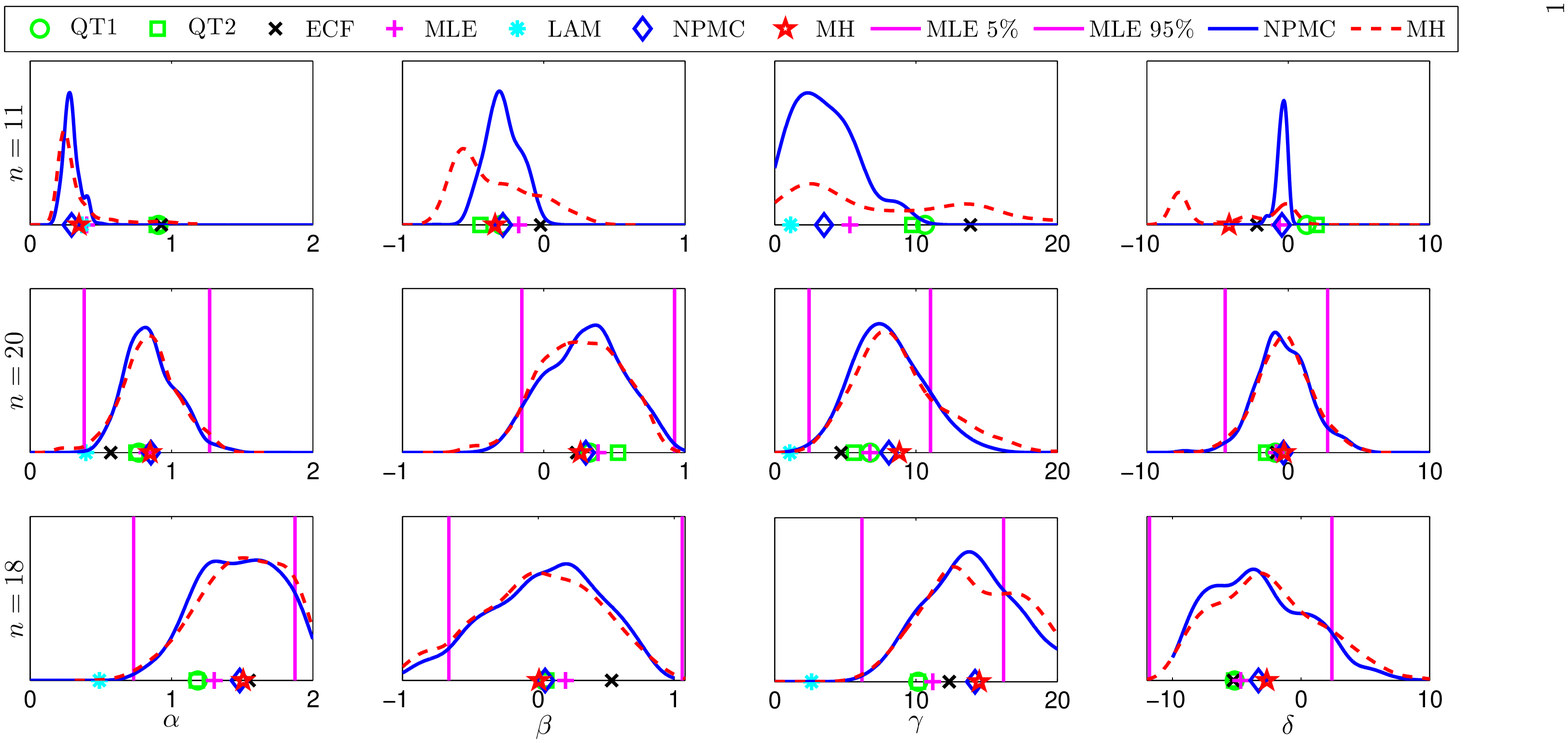}
\vspace{-0.9cm}\caption{Point estimates of the $\alpha$-stable parameters provided by the QT1, QT2, ECF, MLE, LAM, NPMC and MH methods, for $n=11,20,18$. The 5\% and 95\% confidence intervals of MLE and the kernel posterior approximations provided by the NPMC and MH methods are also shown. For $n=11$ the MLE method does not yield confidence intervals.}
\label{fig_9}


\vspace{0.3cm}
\centering
\footnotesize{Table 3: Point estimates of the parameters obtained by each of the methods, for the selected data sets $n=11,20,18$. 95\% confidence intervals are given in parentheses for MLE, NPMC and MH.}
\label{tabla_peces}
\begin{tabular}{cccccccccc}
  \hline
  $n$ & $T_n$ & param & QT1 & QT2 & ECF & MLE & LAM & NPMC & MH \\
	\hline
  \multirow{4}{*}{11} & \multirow{4}{*}{35}  & $\hat \alpha$ & 0.908 & 0.896 & 0.927 & 0.400 (0) &  0.366 & 0.293 (0.112) & 0.345 (0.419)\\
	&  & $\hat \beta$ & -0.300 & -0.447 & -0.022 & -0.177 (0) &  0 & -0.289 (0.228) & -0.344 (0.519)\\
	& & $\hat \gamma$ & 10.647 & 9.739 & 13.856 & 5.323 (0) &  1.136 & 3.496 (4.326) & 20.138 (28.565)\\
	& & $\hat \delta$ & 1.306 & 2.003 & -2.216 & -0.636 (0) & 0 & -0.445 (0.667) &-4.185 (6.942) \\
	\hline
  \multirow{4}{*}{20} & \multirow{4}{*}{25}  & $\hat \alpha$ & 0.768 & 0.752 & 0.570 & 0.826 (0.444) & 0.394 & 0.855 (0.351) & 0.849 (0.408)\\
	&  & $\hat \beta$ & 0.325 & 0.526 & 0.233 & 0.385 (0.540) & 0 & 0.297 (0.533) & 0.260 (0.557) \\
	& & $\hat \gamma$ & 6.761 & 5.610 & 4.697 & 6.736 (4.295) & 1.076 &  8.082 (4.720) & 8.828 (6.218) \\
	& & $\hat \delta$ & -0.914 & -1.563 & -0.819 & -0.838 (3.621) & 0 & -0.339 (3.768) & -0.288 (3.898) \\
	\hline
  \multirow{4}{*}{18} & \multirow{4}{*}{27}  & $\hat \alpha$ & 1.186 & 1.179 & 1.549 & 1.303 (0.570) & 0.490 & 1.483 (0.556) & 1.507 (0.596) \\
	&  & $\hat \beta$ & 0.045 & 0.059 & 0.540 & 0.200 (0.858) & 0 & 0.050 (0.801) & 0.005 (0.869)\\
	& & $\hat \gamma$ & 10.161 & 10.149 & 12.356 & 11.184 (5.011) & 2.617 & 14.187 (6.673) & 14.479 (7.232) \\
	& & $\hat \delta$ & -5.164 & -5.213 & -5.263 &  -4.698 (7.102) & 0 & -3.315 (7.307) & -2.663 (8.081) \\
	\hline
\end{tabular}
\end{figure*}

We have applied the NPMC, the MH and the described frequentist methods to this problem and compared the obtained results. For the Bayesian schemes, we have considered prior marginal distributions $p_3(\gamma) = \mathcal{U}(\gamma; (0,50])$ and $p_3(\delta) = \mathcal{U}(\delta; [-10,10])$. The parameters have again been set to $L=10$, $M=10^3$ and $M_T = 30$ for the NPMC method. In order to have a similar computational complexity, the total number of iterations of the MH method has been set to $I = 10^4$, yielding a final sample of $M=1000$ after removing the burn-in period and thinning.

Figure 8 shows the final NESS obtained by the NPMC (\textit{left}) and the MH (\textit{right}) methods, versus the corresponding values of $\alpha$ estimated by each algorithm, similarly to Figure \ref{fig_2} (\textit{left}) and Figure \ref{fig_3} (\textit{left}) in the computer simulations of Section \ref{Sims}. It is important to note that more than 50\% of the individuals are identified as having values of $\alpha < 0.5$, both with NPMC and MH algorithms, which is also confirmed by the MLE method. The particular cases $n=11,20,18$, whose observations are shown in Figure \ref{fig_7}, are depicted with big markers. It can be observed that similar results are obtained in the real data case, where low $\alpha$ values yield low final NESS. Since the NESS has proved to be a good indicator of the convergence of the NPMC method, and is related to the MSE evolution, it can be expected that in this real data problem the algorithm performs similarly to the example with synthetic data.

In Figure \ref{fig_9} the point estimates of the $\alpha$, $\beta$, $\gamma$ and $\delta$ parameters provided by the QT1, QT2, ECF, MLE, LAM, NPMC and MH methods are represented for the selected individuals $n=11,20,18$. Additionally, a Gaussian kernel approximation of the posterior distribution of each parameter is shown for the NPMC and MH methods, and the 5\% and 95\% confidence intervals for the MLE (except for $n=11$). As expected from the data inspection, the NPMC method identifies the case $n=11$ as having a heavy-tailed distribution, with $\hat{\alpha}$ around 0.3, which is coherent with the LAM results, the other reliable method for estimating low $\alpha$. The MLE method returns a final estimate of $\hat{\alpha} = 0.4$ and suggests via a warning message that the true value is actually lower. In the estimation of $\beta$, the NPMC and the MLE methods provide very similar results. The MH method yields similar $\alpha$ and $\beta$ estimates but with a larger variance. The estimate of $\gamma$ is inaccurate in this case, but again the NPMC, LAM and MLE methods agree in their estimates. The NPMC and the MLE methods provide $\delta$ estimates close to 0. The MH method obtains very inaccurate estimates of $\gamma$ and $\delta$. In the case of $n=20$, all methods agree to identify $\alpha$ as close to 0.8, except for the LAM method, which has shown to be less accurate when $\alpha > 0.5$ in the simulation study of Section \ref{Sims}. The estimates of the rest of parameters by the different methods are also similar. Finally, the last case $n=18$ is identified as a light-tailed and symmetric distribution, close to a non-standard Gaussian. Table 3 summarizes the obtained results.

The consistency among the compared methods confirms that the available real data can be properly described by an $\alpha$-stable distribution, as suggested by the visual data inspection. The numerical results are coherent with those obtained with synthetic data, both in terms of the NESS of NPMC and MH methods, and in terms of the comparison of the solutions provided by different techniques. The NPMC method provides consistent estimates (comparing different runs) of all parameters for all values of $\alpha$, with an extremely low amount of observations. On the contrary, the MH algorithm fails to identify the parameters when $\alpha$ is low, as can be seen in Figure \ref{fig_5}, and is very sensitive to the prior selection.

\section{Conclusions}
\label{Conclusions}

We have addressed the estimation of the parameters of $\alpha$-stable distributions in a Bayesian framework. We have combined the nonlinear population Monte Carlo (NPMC) scheme of \cite{Koblents2013a} with a classical numerical approximation of the $\alpha$-stable pdf \cite{Nolan1997} and studied, analytically, the impact of this approximation on the convergence of the nonlinear importance sampler. Then, we have provided computer simulations with synthetic data comparing the NPMC method with the main methods proposed in the literature for this problem. The NPMC algorithm clearly outperforms the traditional frequentist methods in terms of MSE, at the expense of a higher computation cost. It also yields better results than other Bayesian methods, such as MH or PMC-ABC methods, providing a lower estimation error with a lower computational effort. Additionally, we have applied the studied methods to a fish displacement real dataset, and obtained coherent and satisfactory results. We have shown, by means of computer experiments, that the proposed technique attains a good performance even for small values of $\alpha$ and with an extremely low number of observed data, where many of the existing techniques usually fail to perform adequately.

\clearpage

\section*{Acknowledgements}

E. K. acknowledges the support of \textit{Ministerio de Educaci\'on} of Spain (\textit{Programa de Formaci\'on de Profesorado Universitario}, ref. AP2008-00469). J. M. acknowledges the partial support of {\em Ministerio de Econom\'{\i}a y Competitividad} of Spain (program Consolider-Ingenio 2010 CSD2008-00010 COMONSENS and project COMPREHENSION TEC2012-38883-C02-01) and the Office of Naval Research Global (award no. N62909-15-1-2011). M. A. R. acknowledges the financial support of the Natural Sciences and Engineering Council of Canada, CAPES and FAPERJ, and thanks J. Nolan for providing a free copy of the STABLE software. A. M. S. acknowledges the financial support from CNPq and CAPES-DGU. The authors thank the constructive comments of anonymous reviewers, which have led to a much more complete version of the manuscript.

\appendix

\section{Simulation of univariate $\alpha$-stable random variables \cite{Chambers1976}}
\label{sampling_method}

Let $U$ and $V$ be independent random variables, $U$ uniformly distributed in
the interval $(-\frac{\pi}{2}, \frac{\pi}{2})$ and $V$ exponentially
distributed with mean 1. For any $0 < \alpha \leq 2$ and $-1 \leq
\beta \leq 1$, when $\alpha \neq 1$, define $W = \frac{1}{\alpha}
\arctan \left( \beta \left( \frac{\pi\alpha}{2} \right) \right)$.
Then
\begin{equation*}
Z = \left\{ \begin{array}{ll}
              \frac{\sin (\alpha(W+U))}{[\cos(\alpha W)\cos(U)]^{1/\alpha}} \left[\frac{\cos(\alpha W + (\alpha - 1)U)}{V}\right]^{(1-\alpha)/\alpha}, & \textrm{if} \; \alpha \neq 1\\
              \frac{2}{\pi} \left[ \left( \frac{\pi}{2} + \beta U \right) \tan(U)  - \beta \log \left( \frac{\frac{\pi}{2}V \cos U}{  \frac{\pi}{2} + \beta U  } \right)
              \right], & \textrm{if} \; \alpha = 1
            \end{array}
 \right.
\end{equation*}
has $\alpha$-stable distribution $\mathcal{S}(z; \alpha, \beta, 0,
1)$ \cite{Nolan2013}. To simulate stable random variables $\mathcal{S}(x; \alpha,
\beta, \gamma, \delta)$ with arbitrary scale and location
parameters, the following transformation can be applied \cite{Nolan2013}
\begin{equation*}
X = \left\{ \begin{array}{ll}
              \gamma [Z - \beta \tan(\frac{\pi \alpha}{2})] + \delta & \alpha \neq 1\\
              \gamma Z + \delta & \alpha = 1
            \end{array}
 \right..
\end{equation*}

\section{Metropolis-Hastings algorithm \cite{Hastings1970}}
\label{MCMC_Nolan}

Algorithm \ref{alg:MHalgorithm} displays the MH algorithm used in the computer simulations of Section \ref{MH_section}.

\begin{algorithm}[htpb]
\caption{Metropolis-Hastings algorithm \cite{Hastings1970} with a likelihood approximation \cite{Nolan1997}.} \label{alg:MHalgorithm}
\begin{algorithmic}
\ENSURE ($i = 1$):
\begin{enumerate}
\item Draw the starting point from the prior distribution $\bftheta^{(1)} \sim p(\bftheta)$. 
\end{enumerate}

\REQUIRE ($i = 2, \ldots, I$):
\begin{enumerate}
\item Draw a proposed sample $\bftheta^\star \sim q(\bftheta | \bftheta^{(i-1)}) = \mathcal{TN} (\bftheta; \bftheta^{(i-1)}, \boldsymbol{\Sigma})$ from a truncated Gaussian distribution.

\item With probability
\begin{equation*}
\min \left\{ 1, \frac{\hat{p}(\by | \bftheta^\star)p(\bftheta^\star)}{\hat{p}(\by | \bftheta^{(i-1)}) p(\bftheta^{(i-1)})} \right\}
\end{equation*}
accept the move setting $\bftheta^{(i)} = \bftheta^\star$. Otherwise store the current value $\bftheta^{(i)} = \bftheta^{(i-1)}$. The likelihood approximation $\hat{p}(\by | \bftheta)$ is computed as in \cite{Nolan1997}.
\end{enumerate}
\end{algorithmic}
\end{algorithm}

%

\section{PMC-ABC algorithm \cite{Beaumont2009}}
\label{PMC_ABC}

Algorithm \ref{alg:PMC_ABC_alg} shows an outline of the PMC-ABC method of \cite{Beaumont2009}, which is used in the simulations of Section \ref{ABC_section}.

\begin{algorithm}[htpb]
\caption{PMC-ABC algorithm \cite{Beaumont2009}.} \label{alg:PMC_ABC_alg}
\begin{algorithmic}
\REQUIRE ($\ell = 1,\ldots,L$):
\begin{enumerate}
\item Select a proposal distribution $q_\ell(\bftheta)$:
\begin{itemize}
\item at iteration $\ell = 1$, let $q_1(\bftheta) = p(\bftheta)$,
\item at iterations $\ell = 2, \ldots, L$, select the proposal pdf as a truncated Gaussian pdf $q_\ell(\bftheta) = \mathcal{TN}(\bftheta; \boldsymbol{\mu}_{\ell-1}, \boldsymbol{\Sigma}_{\ell-1})$, with parameters computed as in equation \eqref{eq_mu_sigma} but using standard weights $w_{\ell-1}^{(i)}$.
\end{itemize}
\item For $i=1,\ldots,M$, simulate $\bftheta_\ell^{(i)} \sim q_\ell(\bftheta)$ and $\by_\ell^{(i)} \sim p(\by | \bftheta_\ell^{(i)})$ until $\rho(\by, \by_\ell^{(i)}) \leq \epsilon_\ell$, where $\rho$ denotes the Euclidean distance between the summary statistics of the observations $\by$ and the samples $\by_\ell^{(i)}$.

\item Compute normalized IWs as $w_\ell^{(i)} \propto p(\bftheta_\ell^{(i)}) / q_\ell(\bftheta_\ell^{(i)})$.
\end{enumerate}
\end{algorithmic}
\end{algorithm}

\section{Proof of Theorem \ref{thBasic}} \label{apThBasic}

We consider the approximate integral $(f,\pi^{M,\epsilon})$ first. Since
\begin{equation}
(f,\pi) = \frac{
    (fg,q)
}{
    (g,q)
} \quad \mbox{and} \quad (f,\pi^{M,\epsilon}) = \frac{
    (fg^\epsilon,q^M)
}{
    (g^\epsilon,q^M)
}, \label{eqNormalizada0}
\end{equation}
where $q^M = \frac{1}{M} \sum_{i=1}^M \delta_{\bftheta^{(i)}}$, it is simple to show that
\begin{equation}
(f,\pi^{M,\epsilon}) - (f,\pi) = \frac{
    (fg^\epsilon,q^M) - (fg,q)
}{
    (g,q)
} + (f,\pi^{M,\epsilon}) \frac{
    (g,q) - (g^\epsilon,q^M)
}{
    (g,q)
}. \label{eqDecom}
\end{equation}
However, since $(g,q)=(1,h) = \int I_\sS(\bftheta)h(\bftheta)d\bftheta$ and $(f,\pi^{M,\epsilon}) \le \| f
\|_\infty$, Eq. \eqref{eqDecom} readily yields
\begin{equation}
| (f,\pi^{M,\epsilon}) - (f,\pi) | \le \frac{
    1
}{
    (1,h)
} \left|
    (fg^\epsilon,q^M) - (fg,q)
\right| + \frac{
    \| f \|_\infty
}{
    (1,h)
} \left|
    (g,q) - (g^\epsilon,q^M)
\right|, \label{eqInicial}
\end{equation}
and, therefore, the problem reduces to computing bounds for errors of the form $| (bg^\epsilon,q^M) - (bg,q) |$, where $b \in B(\sS)$.

Choose any $b \in B(\sS)$. A simple triangle inequality yields
\begin{equation}
| (bg^\epsilon,q^M) - (bg,q) | \le | (bg^\epsilon,q^M) - (bg,q^M) | + | (bg,q^M) - (bg,q) |. \label{eqTriang1}
\end{equation}
Since $q^M = \frac{1}{M} \sum_{i=1}^M \delta_{\bftheta^{(i)}}$, for the second term on the right hand side of \eqref{eqTriang1} we can write
\begin{equation}
\mbE\left[
    | (bg,q^M) - (bg,q) |^p
\right] = \mbE\left[
    \left|
        \frac{1}{M} \sum_{i=1}^M Z^{(i)}
    \right|^p
\right], \label{eqTriang2t}
\end{equation}
where the random variables
\begin{equation}
Z^{(i)} = b(\bftheta^{(i)})g(\bftheta^{(i)}) - (bg,q), \quad i=1,
..., M, \nonumber
\end{equation}
are i.i.d. with zero mean (recall the $\bftheta^{(i)}$'s are i.i.d. draws from $q$). Therefore, it is straightforward to show that
\begin{equation}
 \mbE\left[
    \left|
        \frac{1}{M} \sum_{i=1}^M Z^{(i)}
    \right|^p
\right] \le \frac{
    \tilde c^p a^p \| b \|_\infty^p
}{
    M^\frac{p}{2}
}, \label{eqZygmund}
\end{equation}
where $\tilde c$ is a constant independent of $M$ and $q$, and $a$ is the uniform upper bound for the weight function $g$ provided by A.\ref{asBounds_on_g}, also independent of $M$. Combining \eqref{eqZygmund} with \eqref{eqTriang2t} readily yields
\begin{equation}
\| (bg,q^M) - (bg,q) \|_p \le \frac{
    \tilde c a \| b \|_\infty
}{
    \sqrt{M}
}. \label{eq_tri2term}
\end{equation}
The inequality \eqref{eq_tri2term} implies that there exists an a.s. finite random variable $U_{\upsilon,b}>0$ such that
\begin{equation}
| (bg,q^M) - (bg,q) | \le \frac{
    U_{\upsilon,b}
}{
    M^{\frac{1}{2}-\upsilon}
}, \label{eq_tri2term_eps}
\end{equation}
where $0 < \upsilon < \frac{1}{2}$ is an arbitrarily small constant independent of $M$ (see \cite[Lemma 4.1]{Crisan2014particle}).

If we expand the first term on the right hand side of \eqref{eqTriang1} we arrive at
\begin{eqnarray}
\left|
    (bg^\epsilon,q^M) - (bg,q^M)
\right| &=& \left|
    \frac{1}{M} \sum_{i=1}^M b(\bftheta^{(i)}) \left(
        g^\epsilon(\bftheta^{(i)}) - g(\bftheta^{(i)})
    \right)
\right| \nonumber \\
&\le& \frac{\| b \|_\infty}{M} \sum_{i=1}^M \left|
    g^\epsilon(\bftheta^{(i)}) - g(\bftheta^{(i)})
\right|. \label{eqIneq2t}
\end{eqnarray}
However, using assumption A.\ref{asApprox_g} in the inequality \eqref{eqIneq2t} above, we readily obtain
\begin{equation}
\left|
    (bg^\epsilon,q^M) - (bg,q^M)
\right| \le \| b \|_\infty \epsilon. \label{eq_tri1term_eps}
\end{equation}
Taking together \eqref{eqTriang1}, \eqref{eq_tri2term_eps} and
\eqref{eq_tri1term_eps} we arrive at
\begin{equation}
| (bg^\epsilon,q^M) - (bg,q) | \le \| b \|_\infty \epsilon + \frac{
    U_{\upsilon,b}
}{
    M^{\frac{1}{2}-\upsilon}
} \label{eqBasica}
\end{equation}
and it is immediate to combine the inequality \eqref{eqInicial} with the bound in \eqref{eqBasica}. If we choose  $b=f$ in order to control the first term on the right hand side of \eqref{eqInicial}, and $b=1$ in order to control the second term, we readily find that
\begin{equation}
| (f,\pi^{M,\epsilon}) - (f,\pi) | \le \frac{
    W_{f,\upsilon}
}{
    M^{\frac{1}{2}-\upsilon}
} + \frac{2\| f \|_\infty}{(1,h)}\epsilon, \label{eqProof_ineq1}
\end{equation}
where
$$
W_{f,\upsilon} = \frac{U_{\upsilon,f} + U_{\upsilon,1}}{(1,h)} > 0
$$
is an a.s. finite random variable independent of $M$ and $\epsilon$. This yields the inequality \eqref{eq1Basic} in the statement of the Theorem, with $C=2\| f \|_\infty / (1,h) < \infty$.

The proof for inequality \eqref{eq2Basic} is simpler. A triangle
inequality yields
\begin{equation}
| (f,\bar \pi^{M,\epsilon}) - (f,\pi) | \le 
| (f,\bar \pi^{M,\epsilon}) - (f,\pi^{M,\epsilon}) | + | (f,\pi^{M,\epsilon}) - (f,\pi) | \label{eqTriang2}
\end{equation}
and \eqref{eqProof_ineq1} yields a bound for the second term on the right hand side of  \eqref{eqTriang2}. For the first term, we note that
\begin{equation}
(f,\bar \pi^{M,\epsilon}) = \frac{
    ( f[\varphi^M \circ g^\epsilon], q^M )
}{
    ( \varphi^M \circ g^\epsilon, q^M )
}, \label{eqNormalizada}
\end{equation}
where $\circ$ denotes composition, hence $(\varphi^M \circ g^\epsilon)(\bftheta) = \varphi^M( g^\epsilon (\bftheta) )$. If we combine \eqref{eqNormalizada} and the expression for $(f,\pi^{M,\epsilon})$ in \eqref{eqNormalizada0} we obtain, by the same argument leading to \eqref{eqInicial}, that
\begin{eqnarray}
| (f,\bar \pi^{M,\epsilon}) - (f,\pi^{M,\epsilon}) | &\le& \frac{
    | (f[\varphi^M\circ g^\epsilon], q^M) - (fg^\epsilon,q^M) |
}{
    (\varphi^M \circ g^\epsilon, q^M)
} 
+ \frac{
    \| f \|_\infty | (\varphi^M\circ g^\epsilon, q^M) - (g^\epsilon,q^M) |
}{
    (\varphi^M \circ g^\epsilon, q^M)
}  \nonumber \\
&\le& a | (f[\varphi^M\circ g^\epsilon], q^M) - (fg^\epsilon,q^M) | 
+ a \| f \|_\infty | (\varphi^M\circ g^\epsilon, q^M) - (g^\epsilon,q^M) |, \nonumber\\
\label{eqTriang3}
\end{eqnarray}
where the second inequality follows from the definition of the clipping transformation $\varphi^M$ and the bound $g^\epsilon \geq a^{-1}$ in A.\ref{asBounds_on_geps}.

In order to use \eqref{eqTriang3}, we look into errors of the form $|(b[\varphi^M\circ g^\epsilon], q^M) - (bg^\epsilon,q^M)|$ for arbitrary $b \in B(\sS)$. This turns out relatively straightforward since, from the construction of $\varphi^M$,
\begin{equation}
|(b[\varphi^M\circ g^\epsilon], q^M) - (bg^\epsilon,q^M)| =
\left|
    \frac{1}{M}\sum_{r=1}^{M_T} b(\bftheta^{(i_r)}) \left[
        g^\epsilon(\bftheta^{(i_{M_T})}) - g^\epsilon(\bftheta^{(i_r)})
    \right]
\right| \le 2 a \|b\|_\infty \frac{M_T}{M},\nonumber\\
\label{eqFacil}
\end{equation}
where the inequality follows from the bound $g^\epsilon \leq a$ in A.\ref{asBounds_on_geps}. We can plug \eqref{eqFacil} into \eqref{eqTriang3} twice, first choosing $b=f$ and then $b=1$, in order to control the two
terms in the triangle inequality. As a result, we arrive at the {\em deterministic} bound
\begin{equation}
| (f,\bar \pi^{M,\epsilon}) - (f,\pi^{M,\epsilon}) | \le \frac{
    2a^2\|f\|_\infty M_T
}{
    M
} \le \frac{
    2a^2\|f\|_\infty
}{
    \sqrt{M}
}, \label{eqSegunda}
\end{equation}
where the second inequality follows from the assumption $M_T\le\sqrt{M}$ in the statement of the Theorem.

Plugging \eqref{eqSegunda} and \eqref{eqProof_ineq1} into \eqref{eqTriang2} yields
\begin{equation}
| (f,\bar \pi^{M,\epsilon}) - (f,\pi) | \le \frac{
    W_{f,\upsilon} + 2a^2\| f \|_\infty
}{
    M^{\frac{1}{2}-\upsilon}
} + \frac{2\|f\|_\infty}{(1,h)} \epsilon, \label{eqProof_ineq2}
\end{equation}
which reduces to the inequality \eqref{eq2Basic} in the statement of the Theorem, with $\bar W_{f,\upsilon} = W_{f,\upsilon} + 2a^2 \| f \|_\infty > 0$ an a.s. finite random variable and $C=2\|f\|_\infty / (1,h) < \infty$ a constant, both independent of $M$ and $\epsilon$. $\Box$

%
%
%

\bibliographystyle{IEEEbib}
\bibliography{biblio}

\end{document}